\documentclass[12pt]{article}
\usepackage{epsfig}
\usepackage{psfrag}
\usepackage{latexsym}
\usepackage{indentfirst}
\usepackage{fancyhdr}
\usepackage{amssymb}
\usepackage{amsfonts}
\usepackage{cite}
\usepackage{bbold}
\usepackage[footnotesize]{caption2}
\usepackage{graphicx}
\usepackage[center,footnotesize,hang]{subfigure}
\usepackage{url}

\newcommand{\al}{\alpha}
\newcommand{\be}{\beta}

\newcommand{\De}{\Delta}

\newcommand{\La}{\Lambda}

\newcommand{\sig}{\sigma}
\newcommand{\vphi}{\varphi}

\newcommand{\unity}{1\hspace{-0.15cm}1}

\newcommand{\mean}[1]{\langle#1\rangle}

\newcommand{\beq}{\begin{equation}}
\newcommand{\eeq}{\end{equation}}
\newcommand{\bac}{\beq\begin{array}}
\newcommand{\eac}{\end{array}\eeq}
\newcommand{\ba}{\begin{array}}
\newcommand{\ea}{\end{array}}
\newcommand{\bea}{\begin{eqnarray}}
\newcommand{\eea}{\end{eqnarray}}
\newcommand{\beaa}{\begin{eqnarray*}}
\newcommand{\eeaa}{\end{eqnarray*}}

\newcommand{\nn}{\nonumber}

\def\beq{\begin{equation}}
\def\eeq{\end{equation}}
\def\bea{\begin{eqnarray}}
\def\eea{\end{eqnarray}}
\def\bet{\begin{tabular}}
\def\eet{\end{tabular}}
\def\bes{\begin{subequations}\bea}
\def\ees{\eea\end{subequations}}

\def\be{\begin{equation}}
\def\ee{\end{equation}}
\def\bc{\begin{center}}
\def\ec{\end{center}}
\def\bea{\begin{eqnarray}}
\def\eea{\end{eqnarray}}
\def\dd{\displaystyle}
\def\nn{\nonumber}

\catcode`@=11
\def\marginnote#1{}
\newcount\hour
\newcount\minute
\newtoks\amorpm
\hour=\time\divide\hour by60
\minute=\time{\multiply\hour by60 \global\advance\minute by-\hour}
\edef\standardtime{{\ifnum\hour<12 \global\amorpm={am}%
        \else\global\amorpm={pm}\advance\hour by-12 \fi
        \ifnum\hour=0 \hour=12 \fi
        \number\hour:\ifnum\minute<10 0\fi\number\minute\the\amorpm}}
\edef\militarytime{\number\hour:\ifnum\minute<10 0\fi\number\minute}
\def\draftlabel#1{{\@bsphack\if@filesw {\let\thepage\relax
   \xdef\@gtempa{\write\@auxout{\string
      \newlabel{#1}{{\@currentlabel}{\thepage}}}}}\@gtempa
   \if@nobreak \ifvmode\nobreak\fi\fi\fi\@esphack}
        \gdef\@eqnlabel{#1}}
\def\@eqnlabel{}
\def\@vacuum{}
\def\draftmarginnote#1{\marginpar{\raggedright\scriptsize\tt#1}}
\def\draft{\oddsidemargin 0.0truein
        \def\@oddfoot{\sl preliminary draft \hfil
        \rm\thepage\hfil\sl\today\quad\militarytime}
        \let\@evenfoot\@oddfoot \overfullrule 3pt
        \let\label=\draftlabel
        \let\marginnote=\draftmarginnote
   \def\@eqnnum{(\theequation)\rlap{\kern\marginparsep\tt\@eqnlabel}%
\global\let\@eqnlabel\@vacuum}  }
\catcode`@=12
\begin{document}
\begin{titlepage}
\vspace*{-1cm}
\phantom{hep-ph/***}

\hfill{RM3-TH/09-4}
\hfill{CERN-PH-TH/2009-008}
\hfill{DFPD-09/TH/04}

\vskip 2.5cm
\begin{center}
{\Large\bf Revisiting Bimaximal Neutrino Mixing}

\vskip 0.2 cm
{\Large\bf  in a Model with $S_4$ Discrete Symmetry }
\end{center}
\vskip 0.2  cm
\vskip 0.5  cm
\begin{center}
{\large Guido Altarelli}~\footnote{e-mail address: guido.altarelli@cern.ch}
\\
\vskip .1cm
Dipartimento di Fisica `E.~Amaldi', Universit\`a di Roma Tre
\\
INFN, Sezione di Roma Tre, I-00146 Rome, Italy
\\
\vskip .1cm
and
\\
CERN, Department of Physics, Theory Division
\\
CH-1211 Geneva 23, Switzerland
\\

\vskip .2cm
{\large Ferruccio Feruglio}~\footnote{e-mail address: feruglio@pd.infn.it} and
{\large Luca Merlo}~\footnote{e-mail address: merlo@pd.infn.it}
\\
\vskip .1cm
Dipartimento di Fisica `G.~Galilei', Universit\`a di Padova
\\
INFN, Sezione di Padova, Via Marzolo~8, I-35131 Padua, Italy
\\
\end{center}
\vskip 0.7cm
\begin{abstract}
\noindent
In view of the fact that the data on neutrino mixing are still compatible with a situation where Bimaximal mixing is valid in first approximation and it is then corrected by terms of $\mathcal{O}(\lambda_C)$ (with $\lambda_C$ being the Cabibbo angle), arising from the diagonalization of the charged lepton masses, such that $\delta \sin^2{\theta_{12}}\sim \sin{\theta_{13}}\sim \mathcal{O}(\lambda_C)$ while $\delta \sin^2{\theta_{23}}\sim  \mathcal{O}(\lambda_C^2)$, we construct a model based on the discrete group $S_4$ where those properties are naturally realized. The model is supersymmetric in 4-dimensions and the complete flavour group is $S_4\times Z_4\times U(1)_{FN}$, which also allows to reproduce the hierarchy of the charged lepton spectrum. The only fine tuning needed in the model is to reproduce the small observed value of $r=\Delta m^2_{sun}/\Delta m^2_{atm}$. Once the relevant parameters are set to accommodate $r$ then the spectrum of light neutrinos shows a moderate normal hierarchy and is compatible, within large ambiguities, with the constraints from leptogenesis as an explanation of the baryon asymmetry in the Universe.

\end{abstract}
\end{titlepage}
\setcounter{footnote}{0}
\vskip2truecm
%
\section{Introduction}
It is an experimental fact \cite{data,FogliIndication,MaltoniIndication} that within measurement errors
the observed neutrino mixing matrix \cite{review} is compatible with
the so called Tri-Bimaximal (TB) form \cite{hps}. The best measured neutrino mixing angle $\theta_{12}$ is just about 1$\sigma$ below the TB value $\tan^2{\theta_{12}}=1/2$, while the other two angles are well inside the 1$\sigma$ interval \cite{data}. In a series of papers \cite{TBA4,AFextra,AFmodular,AFL} it has been pointed out that a broken flavour symmetry based on the discrete
group $A_4$ appears to be particularly suitable to reproduce this specific mixing pattern as a first approximation. Other
solutions based on alternative discrete or  continuous flavour groups have also been considered \cite{continuous,others,bmm}, but the $A_4$ models have a very economical and attractive structure, e.g. in terms of group representations and of field content \footnote{Recently, in ref. \cite{lam}, the claim was made that, in order to obtain the TB mixing "without fine tuning", the unique finite group must be $S_4$ or a larger group containing $S_4$. For us this claim is not well grounded being based on an abstract mathematical criterium for a natural model. We do not share the definition of ref. \cite{lam}, based on group theory, of a natural model. For us a physical field theory model is natural if the interesting results are obtained from a lagrangian that is the most general given the stated symmetry and the specified representation content for the flavons. For example, in \cite{AFextra,AFmodular,AFL}, the authors obtain from $A_4$ (which is a subgroup of $S_4$) a natural (in our sense) model for the TB mixing by simply not including symmetry breaking flavons transforming like the 1' and the 1" representations of $A_4$ (a restriction not allowed by the rules specified in ref. \cite{lam}). Rather, for naturalness we require that additional physical properties like the hierarchy of charged lepton masses also follow from the assumed symmetry and are not obtained by fine tuning parameters: for this actually $A_4$ can be more effective than $S_4$ because it possesses three different singlet representations 1, 1' and 1''. }.
In most of the models $A_4$ is accompanied by additional flavour symmetries, either discrete like $Z_N$ or continuous like U(1), which are necessary to eliminate unwanted couplings, to ensure the needed vacuum alignment and to reproduce the observed mass hierarchies. Given the set of flavour symmetries and having specified the field content, the non leading corrections to the TB mixing arising from loop effects and higher dimensional operators can be evaluated in a well defined expansion. In the absence of specific dynamical tricks, in a generic model, all the three mixing angles receive corrections of the same order of magnitude. Since the experimentally allowed departures of $\theta_{12}$ from the TB value $\sin^2{\theta_{12}}=1/3$ are small, at most of $\mathcal{O}(\lambda_C^2)$, with $\lambda_C$ the Cabibbo angle, it follows that both $\theta_{13}$ and the deviation of $\theta_{23}$ from the maximal value are expected in these models to also be at most of $\mathcal{O}(\lambda_C^2)$ (note that $\lambda_C$ is a convenient hierarchy parameter not only for quarks but also in the charged lepton sector with $m_\mu/m_\tau \sim0.06 \sim \lambda_C^2$ and $m_e/m_\mu \sim 0.005\sim\lambda_C^{3-4}$). A value of $\theta_{13} \sim \mathcal{O}(\lambda_C^2)$ is within the sensitivity of the experiments which are now in preparation and will take data in the near future. However, the present data do not exclude a larger value for $\theta_{13}$ with $\theta_{13} \sim \mathcal{O}(\lambda_C)$. In fact, two recent analysis of the available data lead to
$\sin^2{\theta_{13}}=0.016\pm0.010$ at 1$\sigma$ \cite{FogliIndication} and $\sin^2{\theta_{13}}=0.010^{+0.016}_{-0.011}$ at 1$\sigma$ \cite{MaltoniIndication}, which are compatible with both options. If experimentally it is found that $\theta_{13}$ is near its present upper bound, this could be interpreted as an indication that the agreement with the TB mixing is accidental. Then a scheme where instead the Bimaximal (BM) mixing is the correct first approximation modified by terms of $\mathcal{O}(\lambda_C)$ could be relevant. This is in line with the well known empirical observation that $\theta_{12}+\lambda_C\sim \pi/4$, a relation known as quark-lepton complementarity \cite{compl}, or similarly $\theta_{12}+\sqrt{m_\mu/m_\tau} \sim \pi/4$. No compelling model leading without parameter fixing to the exact complementary relation has been produced so far. Probably the exact complementarity relation is to be replaced with something like $\theta_{12}+\mathcal{O}(\lambda_C)\sim \pi/4$ (which we could call "weak" complementarity).

In the present paper we use the expertise recently acquired with non Abelian finite flavour groups to construct a model based on the permutation group $S_4$ which naturally leads to the BM mixing at the leading level. We adopt a supersymmetric formulation of the model in 4 space-time dimensions. The complete flavour group is $S_4\times Z_4 \times U(1)_{FN}$. In the lowest approximation, the charged leptons are diagonal and hierarchical and the light neutrino mass matrix, after see-saw, leads to the exact BM mixing. The model is built in such a way that the dominant corrections to the BM mixing, from higher dimensional operators in the superpotential,  only arise from the charged lepton sector at Next-to-the-Leading-Order (NLO)  and naturally inherit $\lambda_C$ as the relevant expansion parameter. As a result the mixing angles deviate from the BM values by terms of  $\mathcal{O}(\lambda_C)$ (at most), and weak complementarity holds. A crucial feature of the model is that only $\theta_{12}$ and $\theta_{13}$ are corrected by terms of $\mathcal{O}(\lambda_C)$ while $\theta_{23}$ is unchanged at this order (which is essential to make the model agree with the present data). We finally discuss the phenomenology of this model and the outlook.

The article is organized as follows. In the next section we recall the general properties of the neutrino mass matrix leading to the BM mixing in the flavour basis.
In section 3 we explain why $S_4$ is a simple candidate group for the task. In section 4 we discuss the structure of the model at the leading order (LO), where
the resulting lepton mixing is exactly the BM. In section 5 we justify the choice of the vacuum assumed in the previous section, by minimizing the scalar
potential of the theory in the supersymmetric limit. In section 6 we analyze the higher-order corrections needed to modify the BM mixing in order to derive an acceptable
lepton mixing pattern. Section 7 contains a discussion of the phenomenological predictions of the model also including the constraints coming from leptogenesis and, finally, section 8 is devoted to our conclusion.

%
%
\section{Bimaximal Mixing}

We start by recalling that, in the basis where charged leptons are diagonal, the most general mass matrix which corresponds to $\theta_{13}=0$ and $\theta_{23}$ maximal is of the form \cite{grimus08}
\begin{equation}
m_{\nu}=\left(\matrix{
x&y&y\cr
y&z&w\cr
y&w&z}\right)\;.
\label{gl}
\end{equation}
Note that this matrix is symmetric under the 2-3 or the $\mu - \tau$ exchange. For $\theta_{13}=0$ there is no CP violation,  and phases are
only of Majorana type. If we disregard them, we can restrict our considerations to real parameters. There are four of them in eq. (\ref{gl}) which correspond to three mass eigenvalues and one remaining mixing angle, $\theta_{12}$. In particular, $\theta_{12}$ is given by:
\be \label{teta12}
\sin^2{2\theta_{12}}=\frac{8y^2}{(x-w-z)^2+8y^2}
\ee
In the BM case $\sin^2{2\theta_{12}}=1$ is also fixed and an additional parameter, for example $w$, can be eliminated, leading to:
\begin{equation}
m_{\nu BM}=\left(\matrix{
x&y&y\cr
y&z&x-z\cr
y&x-z&z}\right)\;,
\label{gl2}
\end{equation}
This mass matrix is exactly diagonalized by the BM pattern, which we can write, in a particular phase convention, as:
\begin{equation}
U_{BM}= \left(\matrix{
\dd\frac{1}{\sqrt 2}&\dd-\frac{1}{\sqrt 2}&0\cr
\dd\frac{1}{2}&\dd\frac{1}{2}&-\dd\frac{1}{\sqrt 2}\cr
\dd\frac{1}{2}&\dd\frac{1}{2}&\dd\frac{1}{\sqrt 2}}\right)\;.
\label{2}
\end{equation}
In the BM scheme $\tan^2{\theta_{12}}= 1$, to be compared with the latest experimental
determination:  $\tan^2{\theta_{12}}= 0.45\pm 0.04$ (at $1\sigma$) \cite{data,FogliIndication,MaltoniIndication}, so that a rather large non leading correction is needed, as already mentioned in the Introduction.

Both the TB and the BM mixing matrices are of a form that suggests that mixing angles are independent of mass ratios (while for quarks and charged leptons relations like $\lambda_C^2\sim m_d/m_s$ or $m_\mu/m_\tau$ are typical). In fact, in the basis where charged lepton masses are
diagonal, the effective neutrino mass matrix in the BM case is given by
\bea
m_{\nu BM}&=&U_{BM}{\tt diag}(m_1,m_2,m_3)U_{BM}^T\nn\\
&&\nn\\
&=&\left[\frac{m_1}{4}\mathcal{M}_1+\frac{m_2}{4}\mathcal{M}_2+\frac{m_3}{2}\mathcal{M}_3\right]\;.
\label{1k}
\eea
where
\be
\mathcal{M}_1=\left(\matrix{
2&\sqrt 2&\sqrt 2\cr
\sqrt 2&1&1\cr
\sqrt 2&1&1}\right),~~~~~
\mathcal{M}_2=\left(\matrix{
2&-\sqrt 2&-\sqrt 2\cr
-\sqrt 2&1&1\cr
-\sqrt 2&1&1}\right),~~~~~
\mathcal{M}_3=\left(\matrix{
0&0&0\cr
0&1&-1\cr
0&-1&1}\right).
\label{4k}
\ee
The eigenvalues of $m_{\nu}$ are $m_1$, $m_2$, $m_3$ with eigenvectors $(\sqrt{2},1,1)/2$, $(-\sqrt{2},1,1)/2$ and $(0,1,-1)/\sqrt{2}$, respectively.   In terms of the parameters $x$, $y$ and $z$ of eq. (\ref{gl2}) we have   $m_1=x+\sqrt{2}y$,  $m_2=x-\sqrt{2}y$, $m_3=2z-x$.
Clearly, all type of hierarchies among neutrino masses can be accommodated. The smallness of the ratio $r=\Delta m^2_{sun}/\Delta m^2_{atm}=(|m_2|^2-|m_1|^2)/\vert |m_3|^2-|m_2|^2\vert$
requires either $\vert xy\vert \ll \vert z^2\vert$ (normal hierarchy) or $\vert x\vert \sim \vert z\vert\ll \vert y\vert$ (inverse hierarchy) or $\vert y\vert \ll \vert x\vert \sim \vert z\vert$ (approximate degeneracy except for $x\sim 2z$).
From the general expression of the eigenvectors one immediately has a confirmation that this mass matrix, independent of the values of $m_i$, leads to the BM mixing matrix.


\section{A convenient presentation of the group $S_4$}

We first present the argument to show that $S_4$, the permutation group of 4 elements, is a good candidate for a flavour symmetry to realize the BM mixing.  The group $S_4$ has 24 transformations and 5 irreducible representations, which are $3$, $3'$, $2$, $1$ and $1'$.

\begin{table}[h]
\begin{center}
\begin{tabular}{|c|c|c|c|c|c|c|}
  \hline
  & n & $\chi_1$ & $\chi_{1^\prime}$ & $\chi_2$ & $\chi_3$ & $\chi_{3^\prime}$ \\
  \hline
  $C_1$ & $1$ & $1$ & $1$ & $2$ & $3$ & $3$  \\
  $C_2$ & $3$ & $1$ & $1$ & 2 & $-1$ & $-1$  \\
  $C_3$ & $8$  & $1$ & $1$ & $-1$ & $0$ & $0$  \\
  $C_4$ & $6$  & $1$ & $-1$ & $0$ & $1$ & $-1$  \\
  $C_5$ & $6$ & $1$ & $-1$ & $0$ & $-1$ & $1$  \\
  \hline
  \end{tabular}
\end{center}
\begin{center}
\begin{minipage}[t]{12cm}
\caption{\label{tab:S4Classes}Character table of $S_4$. $C_i$ are the conjugacy classes, n the number of elements in each class.}
\end{minipage}
\end{center}
\end{table}

The character table of the group is shown in Table \ref{tab:S4Classes} and, in terms of two operators $S$ and $T$ satisfying to
\beq
T^4=S^2=(ST)^3=(TS)^3=1\;,
\eeq
all the 24 $S_4$ transformations can be obtained by taking suitable products. Divided into the 5 equivalence classes (as many as the inequivalent irreducible representations) the transformations are given in Table \ref{tab:S4Elements}. Different presentations of the $S_4$ group have been discussed in the recent literature \cite{bmm,BaseClaudia,BaseMa} and the choice of generators that we adopt in this paper is related to other existing choices by unitary transformations.

\begin{table}[h]
\begin{center}
\begin{tabular}{|c|c|c|}
  \hline
  & Tr & Products  \\
  $C_1$ &3& $1$ \\
  $C_2$ & -1 &  $T^2$, $ST^2S$,$(ST^2)^2$ \\
  $C_3$ & 0 & $ST$, $TS$,  $(ST)^2$, $(TS)^2$, $T^2ST$, $TST^2$, $T^3ST^2$, $T^2ST^3$ \\
  $C_4$ & 1 &  $S$, $T^3ST$, $TST^3$, $T^2ST^2$, $ST^2ST$, $TST^2S$ \\
  $C_5$ & -1 &  $T$,  $T^3$, $ST^2$, $T^2S$, $STS$, $TST$   \\
\hline
  \end{tabular}
\end{center}
\begin{center}
\begin{minipage}[t]{12cm}
\caption{\label{tab:S4Elements}The 24 $S_4$ transformations obtained as products of $S$ and $T$ and divided into the 5 equivalence classes. In the second column the trace, in the representation 3, of all the representative matrices in each class.}
\end{minipage}
\end{center}
\end{table}

Explicit forms of $S$ and $T$ in each of the irreducible representations can be simply obtained. In the representation $1$ we have $T=1$ and $S=1$, while $T=-1$ and $S=-1$ in $1'$. In the representation $2$ we have:
\beq
T=\left(
    \begin{array}{cc}
      1 & 0 \\
      0 & -1 \\
    \end{array}
  \right)\qquad\qquad
S=\frac{1}{2}\left(
    \begin{array}{cc}
      -1 & \sqrt3  \\
      \sqrt3 & 1   \\
    \end{array}
  \right)
  \label{ST2}\;.
\eeq
For the representation $3$, the generators are:
\beq
T=\left(
    \begin{array}{ccc}
      -1 & 0 & 0 \\
      0 & -i & 0 \\
      0 & 0 & i \\
    \end{array}
  \right)\qquad\qquad
S=\left(
    \begin{array}{ccc}
      0 & -\frac{1}{\sqrt{2}} & -\frac{1}{\sqrt{2}}  \\
      -\frac{1}{\sqrt{2}} & \frac{1}{2} & -\frac{1}{2} \\
      -\frac{1}{\sqrt{2}} & -\frac{1}{2} & \frac{1}{2} \\
    \end{array}
  \right)\;.
  \label{matS}
\eeq
In the representation $3'$ the generators $S$ and $T$ are simply opposite in sign with respect to those in the 3 (a glance to Table 2 shows that all elements in the equivalence classes $C_4$ and $C_5$ change sign going from $3$ to $3'$ while those in classes $C_1$, $C_2$ and $C_3$ remain the same).

We now recall the multiplication table for $S_4$ ($R$ stands for any representation). The Clebsch-Gordan coefficients in our basis are collected in the Appendix A.
\[
\begin{array}{l}
1\otimes R=R\otimes1=R\quad\\
1'\otimes1'=1\\
1'\otimes2=2\\
1'\otimes3=3'\\
1'\otimes3'=3\\
\\
2\otimes2=1\oplus1'\oplus2\\
2\otimes3=3\oplus3^\prime\\
2\otimes3'=3\oplus3^\prime\\
\\
3\otimes3=3'\otimes3'=1\oplus2\oplus3\oplus3'\\
3\otimes3'=1'\oplus2\oplus3\oplus3'
\end{array}
\]
This description of the group $S_4$ is particularly suitable for our purposes because the general neutrino mass matrix corresponding to the BM mixing, in the basis where charged leptons are diagonal, given by eq. (\ref{gl2}), is left invariant by the unitary, real and symmetric matrix $S$ in eq. (\ref{matS}):
\beq
m_{\nu BM}= S m_{\nu BM}S\;.
\label{invS}
\eeq
More precisely, the matrix $m_{\nu BM}$ can be completely characterized by the requirement of being invariant under the action of $S$, eqs.~(\ref{matS})-(\ref{invS}), and under the action of $U$, also a unitary, real and symmetric matrix
\beq
m_{\nu BM}= U m_{\nu BM}U\;,~~~~~~~
U=\left(
    \begin{array}{ccc}
      1 & 0 & 0 \\
      0 & 0 & 1 \\
      0 & 1 & 0 \\
    \end{array}
  \right)\;.
\label{invU}
\eeq
The most general solution to the eqs. (\ref{invS}) and  (\ref{invU}), is given by the matrix of eq. (\ref{gl2}).
The matrices $S$ and $U$ commute and generate a discrete group of the type $Z_2\times Z_2$.
As we shall see, in our model the invariance under $U$ arises accidentally, as a consequence of the specific field content
and is limited to the contribution of the dominant terms to the neutrino mass matrix. For this reason we do not need to include the $U$ generator in the flavour symmetry group.


Charged leptons must be diagonal in the basis where $m_{\nu}$ has the BM form. A diagonal matrix $m_l^+m_l$  with generic entries is invariant under $T$ given in eq. \ref{matS}:
\beq
m_l^+m_l= T^+ m_l^+m_l T
\eeq
and, conversely, the most general hermitian matrix invariant under $T$ is diagonal \footnote{This property remains true also when $T$ is replaced by $\eta T$ where $\eta$ represents an arbitrary phase.}. By starting from a flavour symmetry group containing $S_4$, we will realize a special vacuum alignment such that, at LO, the residual symmetry in the neutrino sector will be that generated by $S$ and $U$, while in the charged lepton sector will be, up to a phase, that generated by $T$. Then, by construction, the LO lepton mixing in this model will be of the BM type. Furthermore, a realistic model will be obtained by adding suitable subleading corrections to this zeroth order approximation.

%
\section{The structure of the model}

We discuss here the general properties of our model,  which leads to the BM mixing in first approximation and is formulated in terms of the $S_4$  realization introduced above. We formulate our model in the framework of the see-saw mechanism (even though it would also be possible to build a version where light neutrino masses are directly described by a single set of higher dimensional operators, violating the total lepton number by two units). For this we assign the 3 generations of left-handed (LH) lepton doublets $l$ and of right-handed (RH) neutrinos $\nu^c$ to two triplets $3$, while the RH charged leptons $e^c$, $\mu^c$ and $\tau^c$ transform as $1$, $1'$ and $1$, respectively. The $S_4$ symmetry is then broken by suitable triplet flavons. All the flavon fields are singlets under the Standard Model gauge group.  Additional symmetries are needed, in general, to prevent unwanted couplings and to obtain a natural hierarchy among $m_e$, $m_\mu$ and $m_\tau$. In our model, the complete flavour symmetry is $S_4\times Z_4\times U(1)_{FN}$.  A flavon $\theta$, carrying a negative unit of the $U(1)_{FN}$ charge F, acquires a vacuum expectation value (VEV) and breaks $U(1)_{FN}$. In view of a possible GUT extension of the model at a later stage, we adopt a supersymmetric context, so that two Higgs doublets $h_{u,d}$, invariant under $S_4$, are present in the model. A $U(1)_R$ symmetry related to R-parity and the presence of driving fields in the flavon superpotential are common features of supersymmetric formulations. Supersymmetry also helps producing and maintaining the hierarchy $\langle h_{u,d}\rangle=v_{u,d}\ll \Lambda$ where $\Lambda$ is the cut-off scale of the theory.

The fields in the model and their classification under the symmetry are summarized in Table \ref{table:Transformations}. The complete superpotential can be written as
\beq
w=w_l+w_\nu+w_d\;,
\eeq
\begin{table}[h]
\begin{center}
\begin{tabular}{|c||c|c|c|c|c|c||c||c|c|c|c||c|c|c|c|}
  \hline
  &&&&&&&&&&&&&&&\\[-0,3cm]
  & $l$ & $e^c$ & $\mu^c$ & $\tau^c$ & $\nu^c$ & $h_{u,d}$ & $\theta$ & $\vphi_l$ & $\chi_l$ & $\psi_l^0$ & $\chi_l^0$ & $\xi_\nu$ &$\vphi_\nu$ & $\xi_\nu^0$ & $\vphi_\nu^0$ \\
  &&&&&&&&&&&&&&&\\[-0,3cm]
  \hline
  &&&&&&&&&&&&&&&\\[-0,3cm]
  $S_4$ & 3 & 1 & $1^\prime$ & 1 & 3 & 1 & 1 & 3 & $3^\prime$ & 2 & $3'$ & 1 & 3 & 1 & 3  \\
  &&&&&&&&&&&&&&&\\[-0,3cm]
  $Z_4$ & 1 & -1 & -i & -i & 1 & 1 & 1 & i & i & -1 & -1 & 1 & 1 & 1 & 1 \\
  &&&&&&&&&&&&&&&\\[-0,3cm]
  $U(1)_{FN}$ & 0 & 2 & 1 & 0 & 0 & 0 & -1 & 0 & 0 & 0 & 0 & 0 & 0 & 0 & 0  \\
  &&&&&&&&&&&&&&&\\[-0,3cm]
  $U(1)_R$ & 1 & 1 & 1 & 1 & 1 & 1 & 0 & 0 & 0 & 2 & 2 & 0 & 0 & 2 & 2  \\
  \hline
  \end{tabular}
\end{center}
\begin{center}
\begin{minipage}[t]{12cm}
\caption[]{\label{table:Transformations}Transformation properties of all the fields.}
\end{minipage}
\end{center}
\end{table}
with (we indicate with $(\ldots)$ the singlet 1, with $(\ldots)^\prime$ the singlet $1^\prime$ and with $(\ldots)_R$ ($R=2,\,3,\,3'$) the representation R)
\bea
w_l\;=&&\frac{y_e^{(1)}}{\La^2}\frac{\theta^2}{\La^2}e^c(l\vphi_l\vphi_l)+ \frac{y_e^{(2)}}{\La^2}\frac{\theta^2}{\La^2}e^c(l\chi_l\chi_l)+ \frac{y_e^{(3)}}{\La^2}\frac{\theta^2}{\La^2}e^c(l\vphi_l\chi_l)+\nn\\
&+&\frac{y_\mu}{\La}\frac{\theta}{\La}\mu^c(l\chi_l)^\prime+\frac{y_\tau}{\La}\tau^c(l\vphi_l)+\dots
\label{wl}\\
\nn\\
w_\nu\;=&&y(\nu^cl)+M \Lambda (\nu^c\nu^c)+a(\nu^c\nu^c\xi_\nu)+b(\nu^c\nu^c\vphi_\nu)+\dots\\
\nn\\
w_d\;=&&M_\vphi\Lambda (\vphi_\nu^0\vphi_\nu)+g_1\left(\vphi_\nu^0(\vphi_\nu\vphi_\nu)_3\right)+g_2\left(\vphi_\nu^0\vphi_\nu\right)\xi_\nu+\nn\\
&+&M_\xi^2\La^2\xi_\nu^0+M'_\xi\Lambda\xi_\nu^0\xi_\nu+g_3\xi_\nu^0\xi_\nu\xi_\nu+g_4\xi_\nu^0(\vphi_\nu\vphi_\nu)+\nn\\
&+&f_1\left(\psi_l^0(\vphi_l\vphi_l)_2\right)+f_2\left(\psi_l^0(\chi_l\chi_l)_2\right) +f_3\left(\psi_l^0(\vphi_l\chi_l)_2\right)+\nn\\
&+&f_4\left(\chi_l^0(\vphi_l\chi_l)_{3'}\right)+\dots
\label{wd}\,.
\eea
To keep our formulae compact, we use a two-component notation
for the fermion fields and omit to write the Higgs fields
$h_{u,d}$.  For instance
$y_\tau \tau^c(l\vphi_l)/\La$ stands for $y_\tau \tau^c(l\vphi_l)h_d/\La$,
$y(\nu^cl)$ stands for $y(\nu^cl) h_u$. The powers of the cutoff $\La$ also take  into account the presence of the omitted Higgs fields. Note that the parameters $M$, $M_\phi$, $M_\xi$ and $M'_\xi$ defined above are dimensionless.
In the above expression for the superpotential $w$, only the lowest order operators
in an expansion in powers of $1/\Lambda$ are explicitly shown. Dots stand for higher
dimensional operators that will be discussed later on. The stated symmetries ensure that, for the leading terms, the flavons that appear in $w_l$ cannot contribute to $w_\nu$ and viceversa.

We will show in Sect. 5 that the potential corresponding to $w_d$ possesses an isolated minimum for the following VEV configuration:
\beq
\dd\frac{\mean{\vphi_l}}{\La}=\left(
                     \begin{array}{c}
                       0 \\
                       1 \\
                       0 \\
                     \end{array}
                   \right)A\qquad\qquad
                   \dd\frac{\mean{\chi_l}}{\La}=\left(
                     \begin{array}{c}
                       0 \\
                       0 \\
                       1 \\
                     \end{array}
                   \right)B
\label{vev:charged:best}
\eeq
\beq
\hspace{-1.5cm}
\dd\frac{\mean{\vphi_\nu}}{\La}=\left(
                     \begin{array}{c}
                       0 \\
                       1 \\
                       -1 \\
                     \end{array}
                   \right)C\qquad\quad
\dd\frac{\mean{\xi_\nu}}{\La}=D
\label{vev:neutrinos}
\eeq
where the factors  $A$, $B$, $C$, $D$ should obey to the relations:
\bea
&\sqrt{3}f_1A^2+\sqrt{3}f_2B^2+f_3AB=0
\label{AB}\\
\nn\\
&D=-\dd\frac{M_\vphi}{g_2}\qquad\qquad C^2=\dd\frac{g_2^2M_\xi^2+g_3M_\vphi^2-g_2M_\vphi M'_\xi}{2 g_2^2g_4}
\label{CD}\;.
\eea

Notice the existence of a flat direction related to an arbitrary, common rescaling of $A$ and $B$: if we indicate with $m^2_{\vphi_l}$ and $m^2_{\chi_l}$ the soft masses of the two flavons $\vphi_l$ and $\chi_l$, we can assume $m^2_{\vphi_l},m^2_{\chi_l}<0$ and then $\mean{\vphi_l}$ and $\mean{\chi_l}$ slide to a large scale, which we assume to be eventually stabilized by one-loop radiative corrections, fixing in this way $A$ and $B$.
In the discrete component $S_4\times Z_4$ of the full flavour group  we can choose generators $(S,T,i)$,
where the imaginary unit $i$ denotes the generator of the $Z_4$ factor.
The flavons $\xi_\nu$ and $\varphi_\nu$ are invariant under $Z_4$ and their VEV's are eigenvectors of the generator $S$ corresponding to the eigenvalue 1,
so that the corresponding breaking of $S_4\times Z_4$ preserves the subgroup $G_{\nu}$ generated by $(S,i)$.
In the charged lepton sector $S_4\times Z_4$ is broken down to the subgroup $G_l$, generated by the product $i T$.
Indeed the generator $iT$ is given by $\pm{\tt diag}(-i ,1 ,-1)$, with the plus (minus) sign for the $3~(3')$ $S_4$ representation.
Such a generator, acting in the appropriate representation on the VEV's of eq. (\ref{vev:charged:best}), leaves them invariant.
It is precisely the mismatch, present at the LO, between the subgroups $G_{\nu}$ and $G_l$
preserved in the neutrino and charged lepton sectors, respectively, that produces the BM lepton mixing, as we will explicitly see in this section.

Similarly, the Froggatt-Nielsen flavon $\theta$ gets a VEV, determined by the D-term associated to the local $U(1)_{FN}$ symmetry (see sect 5), and it is denoted by
\beq
\frac{\mean{\theta}}{\La}= t\;.
\label{deft}
\eeq

With this VEV's configuration, the charged lepton mass matrix is diagonal
\beq
m_l=\left(
         \begin{array}{ccc}
           (y_e^{(1)}B^2-y_e^{(2)}A^2+y_e^{(3)}AB)t^2 & 0 & 0 \\
           0 & y_\mu Bt & 0 \\
           0 & 0 & y_\tau A \\
         \end{array}
       \right) v_d
\eeq
so that at LO there is no contribution to the lepton mixing matrix from the diagonalization of charged lepton masses.
In the neutrino sector for the Dirac and RH Majorana matrices we have
\beq
m_\nu^D=\left(
          \begin{array}{ccc}
            1 & 0 & 0 \\
            0 & 0 & 1 \\
            0 & 1 & 0 \\
          \end{array}
        \right)yv_u\qquad\qquad
M_N=\left(
              \begin{array}{ccc}
                2M+2aD & -2bC & -2bC \\
                -2bC & 0 & 2M+2aD \\
                -2bC & 2M+2aD & 0 \\
              \end{array}
            \right)\Lambda\;.
\label{Feq:RHnu:masses}
\eeq
The matrix $M_N$ can be diagonalized by the BM mixing matrix $U_{BM}$, which represents the full lepton mixing at the LO, and the eigenvalues are
\beq
M_1=2|M+aD-\sqrt{2}bC|\Lambda\qquad M_2=2|M+aD+\sqrt{2}bC|\Lambda\qquad M_3=2|M+aD|\Lambda\;.
\eeq
After see-saw, since the Dirac neutrino mass matrix commutes with $M_N$ and its square is a matrix proportional to unity,
the light neutrino Majorana mass matrix, given by the see-saw relation \mbox{$m_\nu=-(m_\nu^D)^TM_N^{-1}m_\nu^D$}, is also diagonalized by the BM mixing matrix and the eigenvalues are
\beq
|m_1|=\frac{|y^2|v_u^2}{2|M+aD-\sqrt{2}bC|}\dd\frac{1}{\Lambda}\qquad
|m_2|=\frac{|y^2|v_u^2}{2|M+aD+\sqrt{2}bC|}\dd\frac{1}{\Lambda}\qquad
|m_3|=\frac{|y^2|v_u^2}{2|M+aD|}\dd\frac{1}{\Lambda}\;.
\label{spec}
\eeq
The light neutrino mass matrix depends on only 2 effective parameters, at LO, indeed the terms $M$ and $aD$ enter the mass matrix in the combination $F\equiv M+a D$. The coefficients $y_e^{(i)}$, $y_\mu$, $y_\tau$, $y$, $a$ and $b$ are all expected to be of $\mathcal{O}(1)$. A priori $M$ could be of $\mathcal{O}(1)$, corresponding to a RH neutrino Majorana mass of $\mathcal{O}(\Lambda)$, but, actually, we will see that it must be of the same order as $C$ and $D$.

We expect a common order of magnitude for the VEV's (scaled by the cutoff $\Lambda$):
\beq
A \sim B \sim v\;,~~~~~~~~~~C \sim D \sim v'\;.
\eeq
However, due to the different minimization conditions that determine $(A,B)$ and $(C,D)$, we may tolerate a moderate hierarchy
between $v$ and $v'$. Similarly the order of magnitude of $t$ is in principle unrelated to those of $v$ and $v'$ (see section 5).
It is possible to estimate the values of $v$ and $t$ by looking at the mass ratios of charged leptons:
\bac{ll}
\left(\dd\frac{m_\mu}{m_\tau}\right)_{exp}\simeq0.06&\qquad\dd\frac{m_\mu}{m_\tau} \sim t\\
\\[-0,3cm]
\left(\dd\frac{m_e}{m_\mu}\right)_{exp}\simeq0.005&\qquad\dd\frac{m_e}{m_\mu} \sim vt\\
\\[-0,3cm]
\left(\dd\frac{m_e}{m_\tau}\right)_{exp}\simeq0.0003&\qquad\dd\frac{m_e}{m_\tau} \sim vt^2\;.
\eac
In order to fit these relations, approximately we must have $t \sim 0.06$ and $v \sim 0.08$ (modulo coefficients of $\mathcal{O}(1)$).\\

To summarize, at LO we have diagonal and hierarchical charged leptons together with the exact BM mixing for neutrinos. It is clear that substantial NLO corrections are needed to bring the model to agree with the data on $\theta_{12}$. A crucial feature of our model is that the neutrino sector flavons  $\vphi_\nu$ and $\xi_\nu$ are invariant under $Z_4$ which is not the case for the charged lepton sector flavons $\vphi_l$ and $\chi_l$. The consequence is that $\vphi_\nu$ and $\xi_\nu$  can contribute at NLO to the corrections in the charged lepton sector, while at NLO $\vphi_l$ and $\chi_l$ cannot modify the neutrino sector couplings. As a results the dominant genuine corrections to the BM mixing only occur  at NLO through the diagonalization of the charged leptons. We will discuss the NLO corrections in section 6 after having proven that the needed VEV alignment is in fact realized at LO.

\subsection{The light neutrino spectrum and the value of r}
\label{SubSection:r}

We now discuss the constraints on the parameters of the neutrino mass matrix in order to get the correct value for the ratio $r=\Delta m^2_{sun}/\Delta m^2_{atm}=(|m_2|^2-|m_1|^2)/\vert |m_3|^2-|m_2|^2 \vert$, which is $r=0.032^{+0.006}_{-0.005}$ at $3\sigma$'s. Like for models based on $A_4$, also in this case some fine tuning is needed to accommodate the value of $r$. In fact the triplet assignment for LH lepton doublets and for RH neutrinos tends to lead to $r\sim1$. We find useful to begin the presentation by analyzing the LO terms, even though a more complete phenomenological discussion
with the inclusion of the NLO contributions will be illustrated in section 7.

We redefine the parameters in eqs. (\ref{spec}) as follows:
\beq
F=M+aD,~~~~~~~~~~~~~Y=-\sqrt{2}bC\;,
\eeq
so that
\beq
Y\sim v'\;.
\eeq
while $F$, like $M$, a priori could be larger, of $\mathcal{O}(1)$.
We make the phases of $F$ and $Y$ explicit by setting
\beq
F\rightarrow Fe^{i\phi_F}\qquad\qquad Y\rightarrow Ye^{i\phi_Y}\;,
\eeq
where now $F$ and $Y$ are real and positive parameters.
Defining the phase difference $\Delta\equiv\phi_Y-\phi_F$ we can explicitly write the absolute values of the neutrino masses as
\bea
|m_1|&=&\frac{1}{\sqrt{F^2+Y^2+2FY\cos{\Delta}}}\frac{|y^2|v_u^2}{2\Lambda}\nonumber \\
|m_2|&=&\frac{1}{\sqrt{F^2+Y^2-2FY\cos{\Delta}}}\frac{|y^2|v_u^2}{2\Lambda}\nonumber\\
|m_3|&=&\frac{1}{F}\frac{|y^2|v_u^2}{2\Lambda}\;.
\label{nuMasses}
\eea
Note that the phase $\Delta$ is related by a non-trivial relation to the Majorana CP phase $2 \al_{21}$, which, by definition, is
the phase difference between the complex eigenvalues $m_1$ and $m_2$ of the neutrino mass matrix.
Furthermore $\cos\Delta$ must be positive in order to guarantee $|m_2|>|m_1|$. By defining $F/Y\equiv f$, we can write the expression of the ratio $r$:
\beq
r=\frac{4f^3\cos{\Delta}}{(f^2+1+2f\cos{\Delta})(1-2f\cos{\Delta})}\;.
\label{eq:rf}
\eeq

In order to have $r$ small either we take $f$ small or $\cos{\Delta}$ small (or both).
If $F\sim \mathcal{O}(1)$ then $f\sim \mathcal{O}(1/v')$, $r\sim 4f\cos{\Delta}$ and $\cos{\Delta}$ must be extremely small : $\cos{\Delta}\sim v' r/4 \sim 10^{-3}$. We prefer to take $f$ small, such that
\beq
r\sim4f^3\cos\Delta\;.
\eeq
If so, in order to accommodate the value of $r$, we only need, for example,  $4\cos\Delta\sim1$ and $f\sim 1/3$. In conclusion, we have to take $F\sim M\sim \mathcal{O}(v')$ and $f=F/Y$ moderately small.
We interpret the relation $M\sim F \sim v'$, needed to reproduce the value of $r$, as related to the fact that the RH neutrino Majorana M mass must empirically be smaller than the cut-off $\Lambda$ of the theory.  In the context of a grand unified theory this corresponds to
the requirement that $M$ is of $\mathcal{O}(M_{GUT})$ rather than of $\mathcal{O}(M_{Planck})$.

With these positions, the neutrino spectrum shows a moderate normal hierarchy, with
\bea
|m_{1,2}|&\sim&\frac{1}{Y}\;\sim\;\mathcal{O}\left(\frac{1}{v'}\right)\\
|m_3|&\sim&\mathcal{O}\left(\frac{1}{fY}\right)\;,
\eea
in units of $|y^2|v_u^2/2\Lambda$. At the leading order an inverted ordering of the neutrino masses is forbidden, as
we can see from eq. (\ref{nuMasses}), which, for $|m_2|>|m_1|$, always implies $|m_3|>|m_1|$.\\

At the LO the lightest neutrino mass $|m_1|$ has a lower bound of order $0.01$ eV. Indeed, the only possible way to decrease $|m_1|$
is to take $Y$ as large as possible. By expanding eqs. (\ref{nuMasses}) in powers of $f$, we have
\be
|m_1|\approx f |m_3|\ge f \sqrt{\Delta m^2_{atm}}\ge 0.01~ {\rm eV}~~~.
\ee
As we will see in section 7, this lower bound can be evaded by including NLO corrections, but values of $|m_1|$ much smaller than $0.01$ eV would require an additional tuning
of the parameters.

\subsection{Effective Terms}

Since the neutrino sector is not charged under the $Z_4$ symmetry, we have operators of dimension 5 which contribute to the neutrino masses and may corresponds to some heavy exchange other than the RH neutrinos $\nu^c$. These effective terms are
\beq
w_\nu^{eff}=\dd\frac{M'}{\La}(lh_ulh_u)+\dd\frac{a'}{\La^2}(lh_ulh_u)\xi_\nu+\dd\frac{b'}{\La^2}(lh_ulh_u\vphi_\nu)\;.
\eeq
The choice of the names $M'$, $a'$ and $b'$ is not accidental, in fact the resulting mass matrix has a flavour structure identical to that one of $M_N$ in eq.(\ref{Feq:RHnu:masses}):
\beq
m_\nu^{eff}=\left(
              \begin{array}{ccc}
                2M'+2a'D & -2b'C & -2b'C \\
                -2b'C & 0 & 2M'+2a'D \\
                -2b'C & 2M'+2a'D & 0 \\
              \end{array}
            \right)\dd\frac{v_u^2}{\Lambda}\;.
\label{meff}
\eeq
As a consequence, also this mass matrix can be diagonalized by the BM pattern. The complete neutrino mass matrix is then given by the sum of two contributions,
\beq
m_\nu=-(m_\nu^D)^TM_N^{-1}m_\nu^D+m_\nu^{eff}\;,
\eeq
and can still be diagonalized by the the BM scheme. Moreover, since both these terms have the same structure, $m_\nu$ can be parameterized by only three parameters: in principle we can absorb the effective contributions in the see-saw ones.

If we consider $y\sim \mathcal{O}(1)$, $F\sim v'$ and $M'\sim \mathcal{O}(1)$ we get that
\bea
&&m_{1,2}={\mathcal O}(\frac{1}{v'}\dd\frac{v_u^2}{\Lambda})+{\mathcal O}(\dd\frac{v_u^2}{\Lambda})\\
&&m_3={\mathcal O}(\frac{1}{fv'}\dd\frac{v_u^2}{\Lambda})+{\mathcal O}(\dd\frac{v_u^2}{\Lambda})\;.
\eea
If also $M'\sim \mathcal{O}(v')$ then the effective terms are even more suppressed. In conclusion the effective terms are subdominant for the interesting domain of parameters.

\section{Vacuum alignment}

In this section we show that the superpotential $w_d$ given in eq. (\ref{wd}) has an
isolated minimum that corresponds to the VEV's in eqs. (\ref{vev:charged:best}) and (\ref{vev:neutrinos}).
At the LO, the equations for the minimum of the potential can be divided into two decoupled parts: one for
the neutrino sector and one for the charged lepton sector. Given this separation, for shorthand we can omit
the indexes $l$ and $\nu$ for flavons and  driving fields. The driving fields are assumed to have vanishing VEV's.
In the neutrino sector, the equations for the vanishing of the derivatives of $w_d$ with respect to each component
of the driving fields are:
\bea
&M_\vphi\Lambda\vphi_1+g_1(\vphi_3^2-\vphi_2^2)+g_2\xi\vphi_1=0\\
&M_\vphi\Lambda\vphi_3-2g_1\vphi_1\vphi_2+g_2\xi\vphi_3=0\\
&M_\vphi\Lambda\vphi_2+2g_1\vphi_1\vphi_3+g_2\xi\vphi_2=0\\
\nn\\
&M_\xi^2\La^2+M'_\xi\Lambda\xi+g_3\xi^2+g_4(\vphi_1^2+2\vphi_2\vphi_3)=0\;.
\eea
A solution to these equations is given in eqs. (\ref{vev:neutrinos}), (\ref{CD}).
The triplet VEV is eigenvector of the operator $S$, which in the 3
representation is given by the matrix in eq. (\ref{matS}), with eigenvalue 1.

For the charged lepton sector the equations are:
\bea
&f_1(\vphi_1^2-\vphi_2\vphi_3)+f_2(\chi_1^2-\chi_2\chi_3)+\dd\frac{\sqrt{3}}{2}f_3(\vphi_2\chi_2+\vphi_3\chi_3)=0\\
&\dd\frac{\sqrt{3}}{2}f_1(\vphi_2^2+\vphi_3^2)+\dd\frac{\sqrt{3}}{2}f_2\left(\chi_2^2+\chi_3^2\right)+f_3\left[-\vphi_1\chi_1+ \dd\frac{1}{2}\left(\vphi_2\chi_3+\vphi_3\chi_2\right)\right]=0\\
\nn\\
&f_4(\vphi_3\chi_3-\vphi_2\chi_2)=0\\
&f_4(-\vphi_1\chi_2-\vphi_2\chi_1)=0\\
&f_4(\vphi_1\chi_3+\vphi_3\chi_1)=0\;.
\label{Feq:wd:charged}
\eea
There are two independent solutions of this set of equations. One is given in eqs. (\ref{vev:charged:best}), (\ref{AB}).
A second solution is given by:
\beq
\dd\frac{\mean{\vphi_l}}{\La}=\left(
                     \begin{array}{c}
                       -\sqrt{2} \\
                       1 \\
                       1 \\
                     \end{array}
                   \right)A\qquad
\qquad
\dd\frac{\mean{\chi_l}}{\La}=\left(
                     \begin{array}{c}
                       \sqrt{2} \\
                       1 \\
                       1 \\
                     \end{array}
                   \right)B
\label{Fvev:charged:wrong}
\eeq
where the factors $A$ and $B$ should obey to the relation
\beq
f_1A^2+f_2B^2+\sqrt{3}f_3AB=0\;.
\label{ABbis}
\eeq
At this level we assume that some additional input neglected so far in the analysis, such as for instance some choice of the
soft supersymmetry breaking parameters, selects the first solution as the the lowest minimum of the scalar potential.

It is important to note that the stability of the alignment in eqs. (\ref{vev:charged:best}) and (\ref{vev:neutrinos}), under small perturbations can be proven.
If one introduces small parameters in the VEV of the fields as follows
\beq
\dd\frac{\mean{\vphi_l}}{\La}=\left(
                     \begin{array}{c}
                       x_1 \\
                       1 \\
                       x_2 \\
                     \end{array}
                   \right)A\qquad
\qquad
\dd\frac{\mean{\chi_l}}{\La}=\left(
                     \begin{array}{c}
                       y_1 \\
                       y_2 \\
                       1 \\
                     \end{array}
                   \right)B\qquad
\qquad
\dd\frac{\mean{\vphi_\nu}}{\La}=\left(
                     \begin{array}{c}
                       z \\
                       1 \\
                       -1 \\
                     \end{array}
                   \right)C\;,
\eeq
it is only a matter of a simple algebra to show that, for small $(z, x_1,x_2,y_1,y_2)$, the only
solution minimizing the scalar potential in the supersymmetric limit is indeed that given by \mbox{$(z, x_1,x_2,y_1,y_2)=(0,0,0,0,0)$}.

Given the symmetry of the superpotential $w_d$, starting from the field configurations of eqs. (\ref{vev:neutrinos})-(\ref{CD}),
(or (\ref{vev:charged:best})-(\ref{AB})) and acting on them with elements of the flavour
symmetry group $S_4\times Z_4$, we can generate other minima of the scalar potential. Some of them are
\beq
\dd\frac{\mean{\vphi_\nu}}{\La}=\left(
                   \begin{array}{c}
                     1 \\
                     0 \\
                     0 \\
                   \end{array}
                 \right)\qquad\qquad
\dd\frac{\mean{\vphi_\nu}}{\La}=\left(
                   \begin{array}{c}
                     0 \\
                     1 \\
                     1 \\
                   \end{array}
                 \right)\;.
\eeq
The new minima are however physically equivalent to those of the original set and it is not restrictive to analyze the model by choosing as local minimum the specific field configuration
discussed so far.\\

For the Froggatt-Nielsen field the non vanishing VEV is determined by the D-term associated with the $U(1)_{FN}$ symmetry.
In fact, we regard this $U(1)_{FN}$ as a local symmetry. Since the field content displayed in Table 3 is anomalous,
we assume the existence of additional heavy multiplets to cancel the anomaly. Here we do not need to specify these fields,
but we must presume that their VEV's are all vanishing. The D-term in the potential is given by:
\beq
V_D=\frac{1}{2}(M^2_{FI}-g_{FN}|\theta|^2+\dots)^2
\eeq
where $g_{FN}$ is the gauge coupling constant of $U(1)_{FN}$ and $M^2_{FI}$ is the Fayet-Iliopoulos term.
The vanishing of $V_D$ requires
\beq
g_{FN}|\theta|^2=M^2_{FI}.
\eeq
Assuming that $M^2_{FI}/g_{FN}$ is positive, this condition fixes the VEV of $\theta$ and the value of $t$,
given in eq. (\ref{deft}). Due to its particular origin the value of $t$ is in principle unrelated to those of $v$ and $v'$.

\section{Next-to-leading corrections}

We now study the set of subleading corrections to the superpotential that are essential to bring the model in agreement with the data. We start with the corrections to $w_d$ that determines the vacuum alignment and then consider the modification to $w_l$ and $w_\nu$. The corrections can be classified according to an expansion in inverse powers of $\Lambda$
and we will analyze the NLO terms to the LO results discussed before.

\subsection{NLO corrections to vacuum alignment}

The superpotential term $w_d$, linear in the driving fields $\psi_l^0$, $\chi_l^0$, $\xi_\nu^0$ and $\vphi_\nu^0$, is modified into:
\beq
w_d=w^0_d+\Delta w_d\;.
\eeq
where $\Delta w_d$ is the NLO contribution, suppressed by one power of $1/\Lambda$ with respect to $w_d$.  The corrective term $\Delta w_d$ is given by the most general quartic, $S_4\times Z_4 \times U(1)_{FN}$-invariant polynomial linear in the driving fields, and can be obtained by inserting an additional flavon field in all the LO terms. The $Z_4$-charges prevent any addition of the flavons $\vphi_l$ and $\chi_l$ at NLO, while a factor of $\xi_\nu$ or $\vphi_\nu$ can be added to all the LO terms. The full expression of $\Delta w_d$ is the following:
\beq
\Delta w_d=\frac{1}{\La}\left(\sum_{i=1}^3x_iI_i^{\xi_\nu^0}+\sum_{i=1}^{5}w_iI_i^{\vphi_\nu^0}+\sum_{i=1}^6s_iI_i^{\psi_l^0}+
\sum_{i=1}^5v_iI_i^{\chi_l^0}\right)
\eeq
where $x_i$, $w_i$, $s_i$ and $v_i$ are coefficients and $\left\{I_i^{\xi_\nu^0},\;I_i^{\vphi_\nu^0},\;I_i^{\psi_l^0},\;I_i^{\chi_l^0}\right\}$ represent a basis of independent quartic invariants:
\bac{ll}
I_1^{\xi_\nu^0}=\xi_\nu^0\xi_\nu\xi_\nu\xi_\nu\qquad\qquad&
I_3^{\xi_\nu^0}=\xi_\nu^0\xi_\nu(\vphi_\nu\vphi_\nu)\\
I_2^{\xi_\nu^0}=\xi_\nu^0(\vphi_\nu(\vphi_\nu\vphi_\nu)_3)\qquad\qquad&
\eac
\bac{ll}
I_1^{\vphi_\nu^0}=(\vphi_\nu^0\vphi_\nu)(\vphi_\nu\vphi_\nu)\qquad\qquad&
I_4^{\vphi_\nu^0}=\left(\vphi_\nu^0(\vphi_\nu\vphi_\nu)_3\right)\xi_\nu\\
I_2^{\vphi_\nu^0}=\left((\vphi_\nu^0\vphi_\nu)_2(\vphi_\nu\vphi_\nu)_2\right)\qquad\qquad& I_5^{\vphi_\nu^0}=(\vphi_\nu^0\vphi_\nu)\xi_\nu\xi_\nu\\
I_3^{\vphi_\nu^0}=\left((\vphi_\nu^0\vphi_\nu)_3(\vphi_\nu\vphi_\nu)_3\right)\qquad\qquad&
\eac
\bac{ll}
I_1^{\psi_l^0}=\left((\psi_l^0\vphi_\nu)_3(\vphi_l\chi_l)_3\right)\qquad\qquad&
I_4^{\psi_l^0}=\left(\psi_l^0(\vphi_l\vphi_l)_2\right)\xi_\nu\\
I_2^{\psi_l^0}=\left((\psi_l^0\vphi_\nu)_{3'}(\vphi_l\chi_l)_{3'}\right)\qquad\qquad&
I_5^{\psi_l^0}=\left(\psi_l^0(\chi_l\chi_l)_2\right)\xi_\nu\\
I_3^{\psi_l^0}=\left((\psi_l^0\vphi_\nu)_3(\chi_l\chi_l)_3\right)\qquad\qquad&
I_6^{\psi_l^0}=\left(\psi_l^0(\vphi_l\chi_l)_2\right)\xi_\nu\\
\eac
\bac{ll}
I_1^{\chi_l^0}=(\chi_l^0\vphi_\nu)'(\vphi_l\chi_l)'\qquad\qquad&
I_4^{\chi_l^0}=\left((\chi_l^0\vphi_\nu)_{3'}(\vphi_l\chi_l)_{3'}\right)\\
I_2^{\chi_l^0}=\left((\chi_l^0\vphi_\nu)_2(\vphi_l\chi_l)_2\right)\qquad\qquad&
I_5^{\chi_l^0}=\left(\chi_l^0(\vphi_l\chi_l)_{3'}\right)\xi_\nu\;.\\
I_3^{\chi_l^0}=\left((\chi_l^0\vphi_\nu)_3(\vphi_l\chi_l)_3\right)\qquad\qquad&
\eac

The new VEV configuration is obtained by imposing the vanishing of the first derivative of $w_d+\Delta w_d$ with respect to the driving fields $\xi_\nu^0$, $\vphi_\nu^0$, $\psi_l^0$ and $\chi_l^0$. We look for a solution that perturbs eqs. (\ref{vev:charged:best}) and (\ref{vev:neutrinos}) to first order in the $1/\La$ expansion: for all components of the flavons $\Phi=(\xi_\nu,~\vphi_\nu, ~\vphi_l, ~\chi_l)$, we denote the shifted VEV's by
\beq
\langle \Phi \rangle=\langle \Phi \rangle_{LO}+\delta \Phi
\eeq
where $\langle \Phi \rangle_{LO}$ are given by eqs. (\ref{vev:charged:best}) and (\ref{vev:neutrinos}).

After some straightforward algebra the results can be described as follows. In the neutrino sector the shifts $\delta \xi_\nu,~\delta \vphi_\nu$ turn out to be proportional to the LO VEV's $\langle \Phi \rangle_{LO}$ and can be absorbed in a redefinition of the parameters $C$ and $D$. Instead, in the charged lepton sector, the shifts $\delta \vphi_l, ~\delta  \chi_l$ have a non trivial structure, so that the LO texture is modified:
\beq
\mean{\vphi_l}=\left(
                     \begin{array}{c}
                       {\delta \vphi_l}_1\\
                       A' \La \\
                       0\\
                     \end{array}
                   \right)\qquad
\qquad\mean{\chi_l}=\left(
                     \begin{array}{c}
                       {\delta \chi_l}_1 \\
                       0 \\
                       B' \La \\
                     \end{array}
                   \right)
\label{vev:charged:nlo}
\eeq
where $A'$ and $B'$ satisfy a relation similar to that in eq. (\ref{AB}) and the shifts ${\delta \vphi_l}_1$ and ${\delta \chi_l}_1$ are proportional to $v'v\La$, that are, in other words, suppressed by a factor $v'$ with respect to the LO entries $A\La$ and $B\La$, respectively.

\subsection{NLO corrections to the mass matrices}

In this section we present the deviations from the LO mass matrices due to higher order operators and to the corrections to the VEV alignment.
The NLO operators contributing to the lepton masses can be obtained by inserting in all possible ways $\xi_\nu$ or $\vphi_\nu$ in the LO operators
and by extracting the $S_4\times Z_4\times U(1)_{FN}$ invariants. Insertions of one power of the flavons $\vphi_l$ or $\chi_l$
are forbidden by the invariance under the $Z_4$ component of the flavour symmetry group.
Now we list the terms that correct the mass matrices.

\subsubsection{Charged Leptons}

The NLO operators contributing to the charged lepton masses originate from two sources: the first one is the introduction of the singlet $\xi_\nu$ in such a way that all the LO operators are multiplied by $\xi_\nu/\La$ and the resulting corrections are diagonal and of relative order $v'$ with respect to the LO results; the second one consists in the insertion of the triplet $\vphi_\nu$ in the LO terms and the NLO superpotential is
\beq
\begin{array}{rl}
    \delta w_l\;=&\dd\frac{\tilde{y}_e^{(1)}}{\La^3}\dd\frac{\theta^2}{\La^2}e^c(l\vphi_l\vphi_l \vphi_\nu)+ \dd\frac{\tilde{y}_e^{(2)}}{\La^3}\dd\frac{\theta^2}{\La^2}e^c(l\chi_l\chi_l \vphi_\nu)+ \dd\frac{\tilde{y}_e^{(3)}}{\La^3}\dd\frac{\theta^2}{\La^2}e^c(l\vphi_l\chi_l \vphi_\nu)+\\[0.3cm]
    &+\dd\frac{\tilde{y}^{(1)}_\mu}{\La^2}\dd\frac{\theta}{\La}\mu^c(l\vphi_l \vphi_\nu)^\prime +\dd\frac{\tilde{y}^{(2)}_\mu}{\La^2}\dd\frac{\theta}{\La}\mu^c(l\chi_l \vphi_\nu)^\prime+\\[0.3cm]
    &+\dd\frac{\tilde{y}^{(1)}_\tau}{\La^2}\tau^c(l\vphi_l \vphi_\nu)+\dd\frac{\tilde{y}^{(2)}_\tau}{\La^2}\tau^c(l\chi_l \vphi_\nu)\;.
\end{array}
\eeq
The charged lepton mass matrix is obtained by adding the contributions of this new set of operators, evaluated with the insertion of the LO VEV's of eqs. (\ref{vev:charged:best})-(\ref{AB}),
to those of the LO superpotential, eq. (\ref{wl}), evaluated with the NLO VEV's of eq. (\ref{vev:charged:nlo}). By omitting all order one coefficients, we find
\beq
m_l=\left(
         \begin{array}{ccc}
           vt^2 & vv't^2 & vv't^2 \\
           v't & t & 0 \\
           v' & 0 & 1 \\
         \end{array}
       \right)vv_d\;.
\eeq
In general each entry receives contributions from both the new operators in $\delta w_l$ and from the LO terms of $w_l$ evaluated with NLO VEVs. Notice however that, at this order, the 23 and 32 elements of $m_l$ are still vanishing.

The unitary matrix that realizes the transformation to the physical basis where the product $(m_l^\dag m_l)$ is diagonal at NLO is of the general form
\beq
U_l=\left(
         \begin{array}{ccc}
           1 & V_{12} v' & V_{13} v' \\
           -V_{12} v' & 1 & 0 \\
           -V_{13} v' & 0 & 1 \\
         \end{array}
       \right)\;,
\eeq
where the coefficients $V_{ij}$ are of $\mathcal{O}(1)$.

\subsubsection{Neutrinos}

The NLO operators contributing to the neutrino masses are given by
\beq
\begin{array}{rl}
   \delta w_\nu\;=&\dd\frac{y_1}{\La}(\nu^cl)\xi_\nu h_u+\dd\frac{y_2}{\La}(\nu^cl\vphi_\nu)h_u+\\[0.3cm]
    &+\dd\frac{a_1}{\La}(\nu^c\nu^c)(\vphi_\nu\vphi_\nu)+\dd\frac{a_2}{\La}((\nu^c\nu^c)_2(\vphi_\nu\vphi_\nu)_2)+\\[0.3cm]
    &+\dd\frac{a_3}{\La}((\nu^c\nu^c)_3(\vphi_\nu\vphi_\nu)_3)+\dd\frac{a_4}{\La}(\nu^c\nu^c\vphi_\nu)\xi_\nu+h.c.\;.
\end{array}
\label{wnu:NLO}
\eeq
As we have already observed, the structure of the LO VEV's of the neutrino flavons is unchanged by the NLO corrections
and the only possible modification to neutrino mass terms originates from the operators listed here.
The Dirac mass matrix becomes
\bea
m_\nu^D=\left(
          \begin{array}{ccc}
            1 & 0 & 0 \\
            0 & 0 & 1 \\
            0 & 1 & 0 \\
          \end{array}
        \right)yv_u+\left(
                      \begin{array}{ccc}
                        y_1D & -y_2C & -y_2C \\
                        -y_2C & 0 & y_1D \\
                        -y_2C & y_1D & 0 \\
                      \end{array}
                    \right)v_u\;.
\label{mnuD:NLO}
\eea
The Majorana mass matrix is also modified by the new operators: we note that the terms proportional to $a_1$ and to $a_4$ can be absorbed by the LO term $aD$ and we refer to all these contributions with the parameter $\hat{a}$; the term proportional to $a_3$ vanishes; the remaining term proportional to $a_2$ fills in new entries of $M_N$ and therefore it is the only genuine correction. The doublet component in the product of two $\vphi_\nu$ has a particular structure and its VEV preserves the subgroup generated by $S$:
\beq
\langle (\vphi_\nu\vphi_\nu)_2\rangle=\left(
                         \begin{array}{c}
                           1 \\
                           \sqrt3 \\
                         \end{array}
                       \right)C^2 \Lambda^2\;.
\eeq
Finally $M_N$ is modified into
\beq
M_N=\left(
              \begin{array}{ccc}
                2M+2\hat{a}D+2a_2C^2 & -2bC & -2bC \\
                -2bC & 3a_2C^2 & 2M+2\hat{a}D-a_2C^2 \\
                -2bC & 2M+2\hat{a}D-a_2C^2 & 3a_2C^2 \\
              \end{array}
            \right)\La\;.
\label{MNNLO}
\eeq
As we can easily see both $M_N$ and $m_\nu$ can be exactly diagonalized by the BM mixing, which represents the total contribution to lepton mixing of the neutrino sector, even after the inclusion of the NLO corrections. These corrections introduce also some terms in the mass eigenvalues of relative order $v'$ with respect to the LO results, but they do not affect the type of the spectrum.\\

Since the neutrino mass matrix is diagonalized by $U_{BM}$, the PMNS matrix can be written as
\beq
U=U_l^\dag U_{BM}\;,
\eeq
and therefore the corrections from $U_l$ affect the neutrino mixing angles at NLO according to
\bac{l}
\sin^2\theta_{12}=\dd\frac{1}{2}-\frac{1}{\sqrt{2}}(V_{12}+V_{13})v'\\[0.2cm]
\sin^2\theta_{23}=\dd\frac{1}{2}\\[0.2cm]
\sin\theta_{13}=\dd\frac{1}{\sqrt{2}}(V_{12}-V_{13})v'\;.
\label{sinNLO}
\eac

\begin{table}[h]
\begin{center}
\begin{tabular}{|c|c|c|}
\hline
&&\\[-4mm]
  & ref. \cite{FogliIndication} & ref. \cite{MaltoniIndication}   \\[2mm]
\hline
&&\\[-4mm]
$\sin^2\theta_{12}$ &$0.326^{+0.050}_{-0.040}~~~[2\sigma]$ & $0.304^{+0.022}_{-0.016}$\\[2mm]
\hline
&&\\[-4mm]
$\sin^2\theta_{23}$ &$0.45^{+0.16}_{-0.09}~~~[2\sigma]$ &  $0.50^{+0.07}_{-0.06}$\\[2mm]
\hline
&&\\[-4mm]
$\sin^2\theta_{13}$ &$0.016\pm0.010$ &$0.010^{+0.016}_{-0.011}$ \\[2mm]
\hline
  \end{tabular}
\end{center}
\begin{center}
\begin{minipage}[t]{12cm}
\caption{\label{table:OscillationData}Results of two recent fits to the lepton mixing angles.}
\end{minipage}
\end{center}
\end{table}

By comparing these expressions with the current experimental values of the mixing angles in table \ref{table:OscillationData}, we see that, to correctly reproduce $\theta_{12}$ we need a parameter $v'$ of the order of the Cabibbo angle $\lambda_C$. Moreover, barring cancellations of/among some the $V_{ij}$ coefficients, also the reactor angle is corrected by a similar amount.

\begin{figure}[h!]
 \centering
   {\includegraphics[width=11cm]{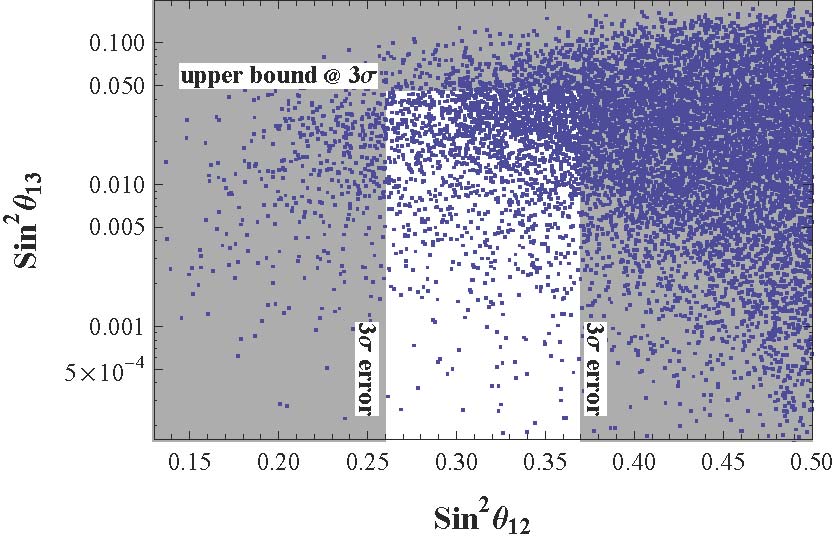}}
 \caption{$\sin^2\theta_{13}$ as a function of $\sin^2\theta_{12}$ is plotted, following eqs. (\ref{sinNLO}). The plot is symmetric with respect $\sin^2\theta_{12}=0.5$ and we report here only the left-part. The parameters $V_{ij}$ are treated as random complex numbers of absolute value between 0 and 2, while $|v'|$ has been fixed at the indicative value of 0.15. The gray bands represents the regions excluded by the experimental data \cite{FogliIndication}: the horizontal one corresponds to the $3\sigma$-upper bound for $\sin^2\theta_{13}$ of 0.46 and the vertical ones to the region outside the $3\sigma$ error range $[0.26 - 0.37]$ for $\sin^2\theta_{12}$.}
 \label{fig0}
\end{figure}

Any quantitative estimates are clearly affected by large uncertainties due to the presence of unknown parameters of order one, as we can see in figure \ref{fig0}, but in our model a value of $\theta_{13}$ much smaller than the present upper bound would be unnatural.

All this discussion works at the NLO, but we expect that at the NNLO the value of $\theta_{23}$ will also be modified with deviations of about $\mathcal{O}(\lambda^2_C)$  at most. The next generation of experiments, in particular those exploiting a high intensity neutrino beam, will probably reduce the experimental error on $\theta_{23}$ and the sensitivity on $\theta_{13}$ to few degrees. If no significant deviations from zero of $\theta_{13}$ will be detected, our construction will be ruled out.

A salient feature of our model is that, at NLO accuracy, the large corrections of $\mathcal{O}(\lambda_C)$ only apply to $\theta_{12}$ and $\theta_{13}$ while $\theta_{23}$ is unchanged at this order. As a correction of $\mathcal{O}(\lambda_C)$ to $\theta_{23}$ is hardly compatible with the present data (see Table 4) this feature is very crucial for the phenomenological success of our model. It is easy to see that this essential property depends on the selection in the neutrino sector of flavons $\xi_\nu$ and $\vphi_\nu$ that transform as 1 and 3 of $S_4$, respectively. We have checked that if, for example, the singlet $\xi_\nu$ is replaced by a doublet $\psi_\nu$ (and correspondingly the singlet driving field  $\xi_\nu^0$ is replaced by a doublet $\psi_\nu^0$), all other quantum numbers being the same, one can construct a variant of the model along similar lines, but in this case all the 3 mixing angles are corrected by terms of the same order. This confirms that a  particular set of $S_4$ breaking flavons is needed in order to preserve $\theta_{23}$ from taking as large corrections as the other two mixing angles.

\section{Phenomenological implications}

We now develop a number of important physical consequences of our model and derive some predictions. We first consider the predicted spectrum and the effective mass $m_{ee}$ for neutrinoless double beta decay and then we discuss the constraints from leptogenesis.

\subsection{Light neutrino masses and neutrinoless double beta decay}
The light neutrino mass matrix, including the NLO corrections described in sections (4.2) and (6.2.2) is given by:
\beq
m_\nu=-(m_\nu^D)^TM_N^{-1}m_\nu^D+m_\nu^{eff}\;,
\eeq
where $m_\nu$, $M_N$ and $m_\nu^{eff}$ are given in eqs. (\ref{mnuD:NLO}), (\ref{MNNLO}), (\ref{meff}), respectively. It is diagonalized
by the BM unitary transformation and its complex eigenvalues are given by:
\bea
m_1&=&\left[2(F'+Y')-\dd\frac{(y'+Y_2)^2}{2(F+Y+a_2 C^2)}\right]\dd\frac{v_u^2}{\Lambda}\nonumber \\
m_2&=&\left[2(F'-Y')-\dd\frac{(y'-Y_2)^2}{2(F-Y+a_2 C^2)}\right]\dd\frac{v_u^2}{\Lambda}\\
m_3&=&\left[-2 F'+\dd\frac{y'^2}{2(F-2 a_2 C^2)}\right]\dd\frac{v_u^2}{\Lambda}\;,\nonumber
\label{numNLO}
\eea
where
\be
F'=M'+a' D~~~,~~~~~Y'=-\sqrt{2} b' C~~~,~~~~~Y_2=-\sqrt{2} y_2 C~~~,~~~~~y'=y+y_1 D~~~.
\ee
We see that the LO expressions are recovered in the limit $F'=Y'=y_{1,2}=0$.
By exploiting eqs. (\ref{numNLO}) we can study some observables like the effective $0\nu2\beta$-decay mass, $|m_{ee}|$, the lightest neutrino mass, $m_1$, and the sum of the neutrino masses directly from the experimental data. We perform a numerical analysis, by treating all the LO, NLO and effective parameters as random complex numbers with absolute value between 0 and 3.
\begin{figure}[h!]
 \centering
 \subfigure[$|m_{ee}|$ vs. $m_1$]
   {\includegraphics[width=7.5cm]{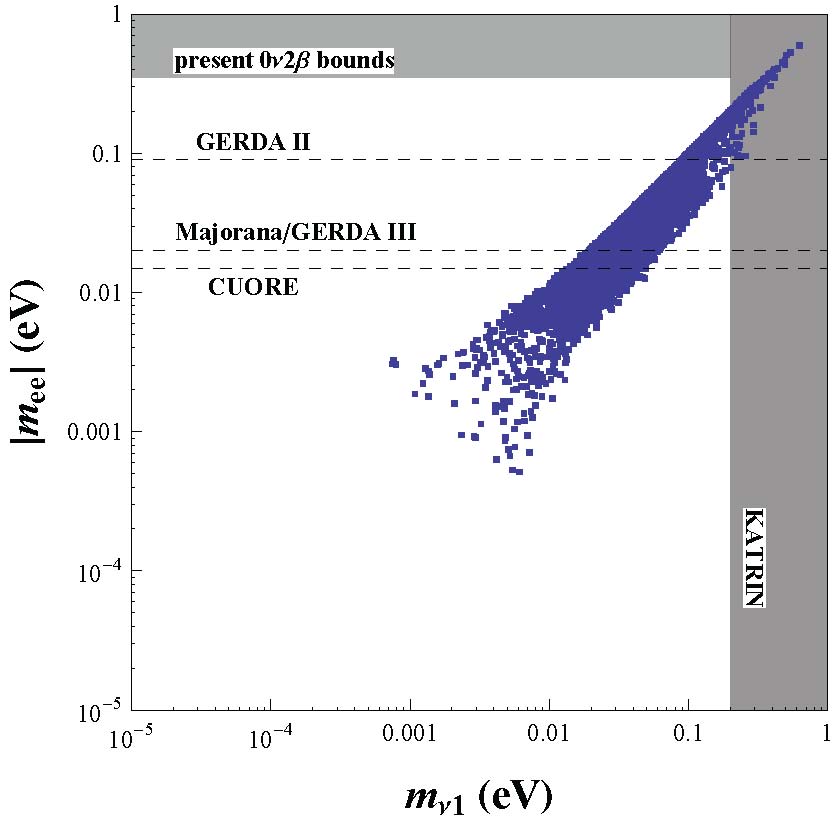}}
 \subfigure[$|m_{ee}|$ vs. $\alpha_{21}$]
   {\includegraphics[width=7.5cm]{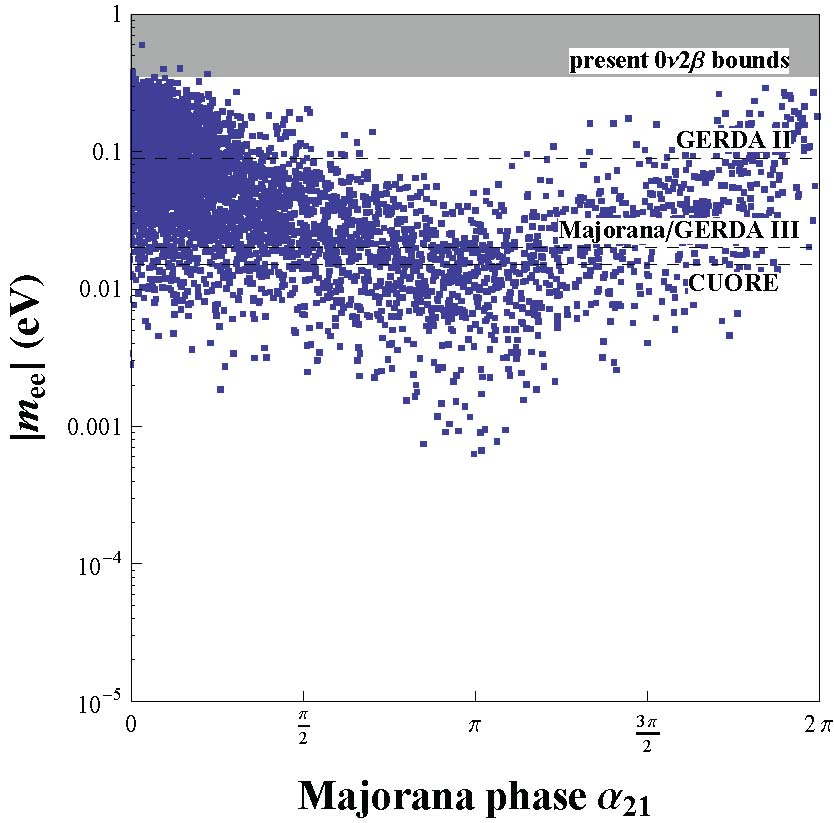}}
 \caption{In figure (a), $|m_{ee}|$ as a function of the lightest neutrino mass, $m_1$, is plotted. The constraints which have been imposed in the plot are the experimental values at $3\sigma$ for $\De m^2_{atm}$, $\De m^2_{sol}$ and the mixing angles. All the parameters of the model are treated as random complex parameters. The present bound from the Heidelberg-Moscow experiment is shown in dark gray and the future sensitivity of CUORE ($\sim15$ meV), Majorana and GERDA III ($\sim20$ meV), and GERDA II ($\sim90$ meV) experiments are represented by the horizontal dashed lines, while the future sensitivity of $0.2$ eV of KATRIN experiment is shown by the vertical band. In figure (b), $|m_{ee}|$ as a function of the physical Majorana phase $\alpha_{21}$ is shown. The experimental bounds are the same as in figure (a).}
 \label{fig1}
\end{figure}
In figure \ref{fig1}(a), we plot $|m_{ee}|$ as a function of the lightest neutrino mass. The points displayed the figures correspond to the case of normal ordering of the neutrino masses, with a moderate hierarchy or a quasi degenerate spectrum.
However, at variance with the results of the leading order, some solutions of our numerical simulation also reproduce an inverted hierarchy
spectrum. The plot displays only the points corresponding to choices of the parameters reproducing
$\De m^2_{atm}$, $\De m^2_{sol}$ and the mixing angles within a 3$\sig$ interval. In dark gray, we plot the present bound from the Heidelberg-Moscow \cite{HM} experiment on $|m_{ee}|$. The vertical band corresponds to the future sensitivity on the lightest neutrino mass of $0.2$ eV from the KATRIN experiment\cite{katrin} and the horizontal lines to the future sensitivity of some $0\nu2\beta$-decay experiments, that are $15$ meV, $20$ meV and $90$ meV, respectively of CUORE \cite{cuore}, Majorana \cite{majorana}/GERDA III\cite{gerda} and GERDA II experiments.
The figure suggests that a lower bound of about $10^{-3}$ eV holds for the lightest neutrino mass. Indeed, with the inclusion of the NLO corrections, from eq.(\ref{numNLO}) we see that $m_{\nu1}$ can vanish if a cancellation between the NLO and LO contributions takes place. This however requires an additional fine tuning of the parameters which has been reproduced in our numerical analysis only partially and by very few points. Similarly the scatter plot indicates a lower bound for $|m_{ee}|$ of about $0.1~meV$. In figure \ref{fig1}(b), we plot $|m_{ee}|$ as a function of the physical Majorana phase. The maximum corresponds to a vanishing value of $\alpha_{21}$ while the minimum to $\alpha_{21}=\pm\pi$, as we expect from the general definition of $|m_{ee}|$. In a non-negligible portion of the parameter space of the model, the predictions for $|m_{ee}|$ approach the future experimental sensitivity. Beyond the experiments already quoted we also recall SuperNEMO ($50\textrm{ meV}$) \cite{supernemo} and EXO ($24\textrm{ meV}$) \cite{exo}.

\begin{figure}[h!]
 \centering
   {\includegraphics[width=8cm]{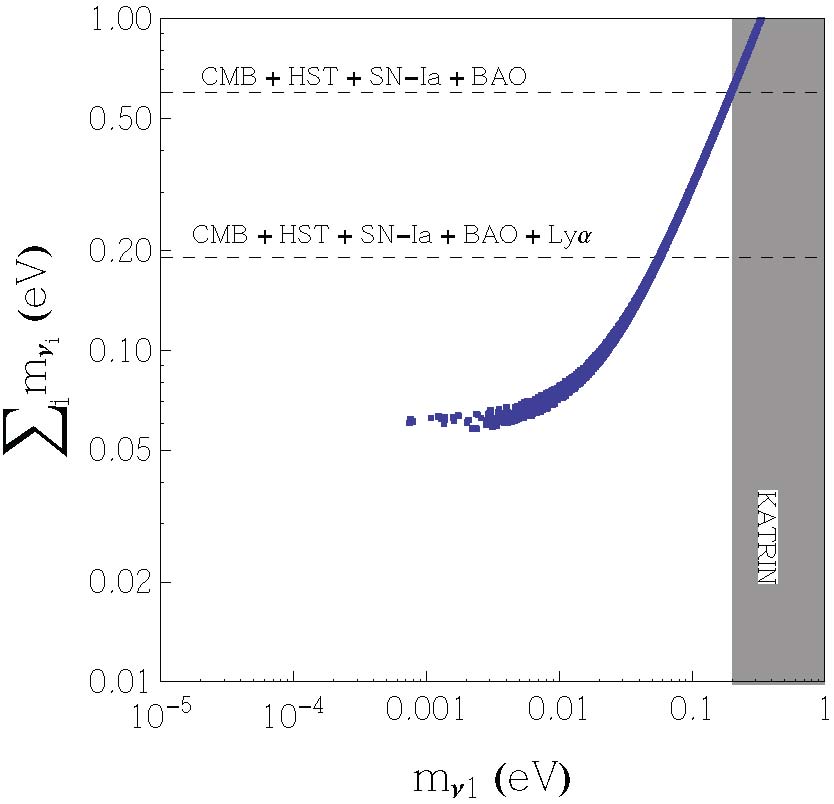}}
 \caption{The sum of the neutrino masses as a function of the lightest neutrino mass $m_1$. The vertical dashed line represents the future sensitivity of $0.2$ eV from KATRIN experiment and the horizontal ones refer to the cosmological bounds
 (see the text for more details).}
 \label{fig2}
\end{figure}

In figure \ref{fig2}, we plot the sum of the neutrino masses as a
function of the lightest neutrino mass, $m_1$. The vertical band refers to the future sensitivity of KATRIN experiment, while the horizontal ones to the cosmological bounds \cite{Cosmology}. There are typically five representative combinations of the cosmological data, which lead to increasingly stronger upper bounds on the sum of the neutrino masses: we are showing the two strongest ones. The first one at $0.60$ eV corresponds to the combination of the Cosmic Microwave Background (CMB) anisotropy data (from WMAP~5y \cite{WMAP2}, Arcminute Cosmology Bolometer Array Receiver (ACBAR) \cite{acbar07}, Very Small Array (VSA) \cite {vsa}, Cosmic Background Imager (CBI) \cite{cbi} and BOOMERANG \cite{boom03} experiments) plus the large-scale structure (LSS) information on galaxy clustering (from the Luminous Red Galaxies Sloan Digital Sky Survey (SDSS) \cite{Tegmark}) plus the Hubble Space Telescope (HST) plus the luminosity distance SN-Ia data of \cite{astier} and finally plus the BAO data from \cite{bao}. The second line at $0.19$ eV corresponds to all the previous data combined to the small scale primordial spectrum from Lyman-alpha (Ly$\alpha$) forest clouds \cite{Ly1}. This plot is typical for normal hierarchy or quasi degenerate spectrum. The only special feature is
the lower bound on $|m_1|$, which, as explained above, relies on a naturalness assumption.

\subsection{Leptogenesis}

In this section we briefly comment on the constraints on our model from the requirement that it is possible to explain a sufficiently large baryon asymmetry through leptogenesis. The baryon asymmetry (see \cite{Davidson} for details) can be defined with respect the entropy of the universe as
\beq
Y_{\Delta B}\equiv \dd\frac{n_B-n_{\overline{B}}}{s}\Big|_0=(8.75\pm0.23)\times10^{-11}\;,
\eeq
where $n_B$, $n_{\overline{B}}$ and $s$ are the number densities of, respectively, baryons, antibaryons and entropy, a subscript $0$ implies "at present time'', and the numerical value is from combined microwave background and large scale structure data. When considering that the baryon asymmetry originates from leptogenesis, it is possible to link $Y_{\Delta B}$ and the lepton asymmetry parameter $\epsilon_i$ as follows:
\beq
Y_{\Delta B}\simeq C\sum_i\epsilon_i\;\eta_i\;,
\eeq
where $C$ represents constants and $\eta_i$ are the efficiency factors linked to the thermal equilibrium of the interactions of the RH neutrinos.
We will not perform here a complete analysis of leptogenesis and in particular we will not evaluate the efficiency factors that, due to washout processes, reduce the final amount of lepton asymmetry. Rather, we formulate our constraint by requiring that the asymmetry parameters $\epsilon_i$ are larger than about $10^{-7}$ (considering a quite usual value for the efficiency factor of about $0.1$ and the supersymmetric context), by assuming that this value, after inclusion of the appropriate conversion factor, reproduces the observed baryon asymmetry. Since in our model RH neutrinos are typically very heavy, with masses above $10^{12}$ GeV, we also assume that the correct framework to analyze leptogenesis is the unflavoured one.

Out-of-equilibrium decays in the early universe of $\nu^c$ to lepton and Higgs doublets produce lepton asymmetries. In a basis in which the Majorana mass matrix for the RH neutrinos is diagonal and real, and where all basis-dependent matrices will be denoted by a hat, the lepton asymmetry parameters are
\beq
\epsilon_i=-\dd\frac{1}{2\pi\left(\hat{Y}_\nu \hat{Y}_\nu^\dag\right)_{ii}}\sum_{j\neq i}\rm{Im}\left\{\left[\left(\hat{Y}_\nu \hat{Y}_\nu^\dag\right)_{ij}\right]^2\right\} g \left(\dd\frac{|M_j|^2}{|M_i|^2}\right)
\eeq
where $\hat{Y}_\nu\equiv \hat{m}_\nu^D/v_u$. Non-vanishing asymmetry parameters require that the off-diagonal entries of the product $\hat{Y}_\nu \hat{Y}_\nu^\dag$ are complex and different from zero, while, at the LO, from eq. (\ref{Feq:RHnu:masses}), we have
\beq
\hat{Y}_\nu=U^T_{BM} Y_\nu\;,~~~~~~~~~~~~~~~~
\hat{Y}_\nu \hat{Y}_\nu^\dag=|y|^2\unity\;,
\eeq
concluding that all the $\epsilon_i$ vanish at LO.

At the NLO the asymmetry parameters still vanish. Indeed, as we have seen in the previous chapter, at NLO both $M_N$ and $Y_\nu$ are still diagonalized by $U_{BM}$. A misalignment between $M_N$ and $Y_\nu Y_\nu^\dagger$ first occurs at the NNLO, when the VEV of the flavon field $\vphi_\nu$ acquires term of order $v^4/v'$ that do not preserve any more the symmetry generated by $S$. The first non-trivial contribution to the off-diagonal entries of $\hat{Y}_\nu \hat{Y}_\nu^\dag$ is of order $v^4/v'$ and arises from the terms proportional to $y_2$ of the superpotential in eq. (\ref{wnu:NLO}), when we take NNLO VEV's of the flavon $\vphi_\nu$. In this case, considering that the lightest RH neutrino is $\nu_3^c$, leptogenesis in governed by $\epsilon_3$: taking at this stage $g\approx 1$, we get
\beq
\epsilon_3\approx \frac{1}{2\pi}\frac{v^8}{v'^2}\;.
\label{eps}
\eeq
Considering the explicit expression of $g$ we get a slightly different result. In this case the RH neutrinos are hierarchical and
the function $g$ scales as
\beq
g\left(\dd\frac{|M_{1,2}|^2}{|M_3|^2}\right)\approx3\dd\frac{|M_3|}{|M_{1,2}|}\;.
\eeq
We can introduce this result in the general expression for $\epsilon_3$ and we have
\beq
\epsilon_3\sim-\frac{3}{2\pi}\frac{v^8}{v'^2}\sum_{i=1,2}\dd\frac{|M_3|}{|M_i|}\;,
\label{epsen}
\eeq
where $|M_3|/|M_{1,2}|\sim1/5$.

\begin{figure}[h!]
\begin{center}
 {\includegraphics[width=8cm]{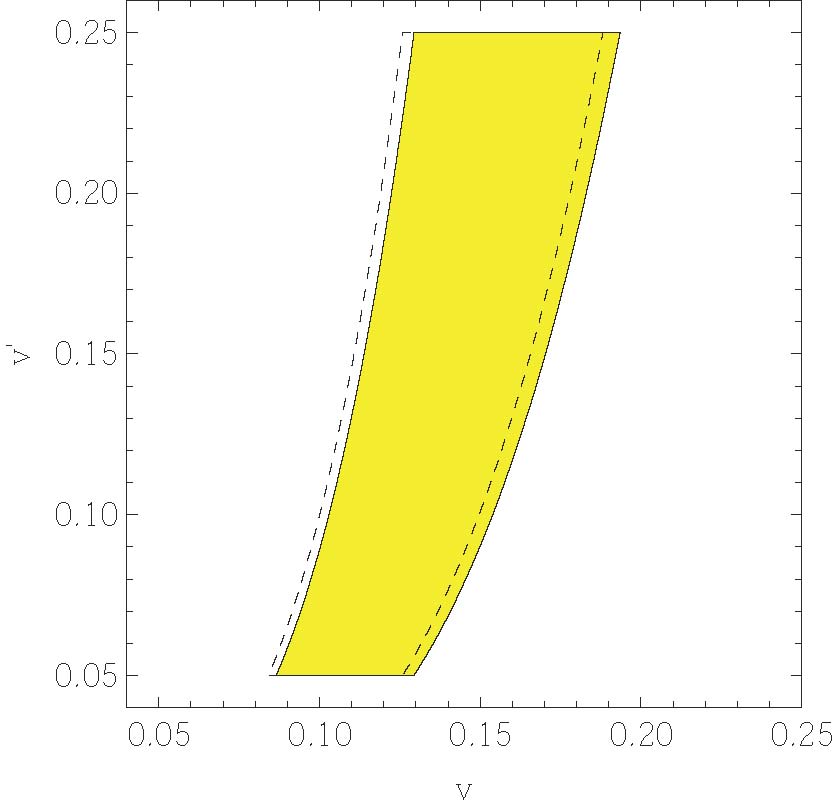}}
\end{center}
\caption{Contour lines of the asymmetry parameter, eq. (\ref{eps}) (continuous lines) and eq. (\ref{epsen}) (dashed lines), in the plane $(v,v')$, for $\epsilon_3=(0.2,5)\times 10^{-6}$.}
\label{BR}
\end{figure}

In figure \ref{BR} we show the region of the parameter plane $(v,v')$ compatible with the requirement $\epsilon_3\approx (0.2 \div 5)\times 10^{-6}$. Clearly, due to the scaling $\epsilon_3\propto v^8$, the most sensitive parameter is $v$, which is bounded to a rather limited region and cannot be too small.
Moreover, even a very large theoretical uncertainty, affecting our choice of the range $\epsilon_3\approx (0.2 \div 5)\times 10^{-6}$, is considerably reduced when translated in terms of the parameter $v$. It is worth to stress that the region of $v$ preferred by leptogenesis has a non-vanishing overlap with the values of $v$ needed to explain the hierarchy in the charged lepton sector.

\section{Summary and conclusion}

We have constructed a see-saw model based on the flavour symmetry $S_4\times Z_4 \times U(1)_{FN}$ where the BM mixing is realized at the LO in a natural way.  The hierarchy of charged lepton masses is obtained as a combined effect of the $U(1)_{FN}$ symmetry breaking measured by the parameter $t$ and of the $S_4\times Z_4$ breaking induced by $v$, proportional to the VEV's of $\vphi_l$ and $\chi_l$. We have $m_\mu/m_\tau =t \sim 0.06$ and  $m_e m_\tau/m_\mu^2 =v \sim 0.08$. Within large uncertainties, a value $v \approx (0.1\div 0.2)$, not too far from the previous figure, is indicated by requiring that the observed baryon asymmetry in the Universe is explained by the amount of leptogenesis predicted by the model.

Since exact BM mixing implies a value of $\tan{\theta_{12}}$ which is excluded by the data, large corrections are needed. The dominant corrections to the BM mixing arise at NLO only through the diagonalization of the charged lepton mass matrix. The shifts of the quantities $\sin^2{\theta_{12}}$ and $\sin{\theta_{13}}$ from the BM values are linear in the parameter $v'$, proportional to the VEV's of $\vphi_\nu$ and $\xi_\nu$, which is expected to be of the same order as $v$, but not necessarily too close, as $v$ and $v'$ are determined by two different sets of minimization equations. From the experimental value $\tan^2{\theta_{12}}= 0.45\pm 0.04$, which is sizeably different than the BM value $\tan^2{\theta_{12}}= 1$, we need $v'\sim \mathcal{O}(\lambda_C)$ where $\lambda_C\sim 0.23$ is the Cabibbo angle.
As in most models where the BM mixing is only corrected by the effect of charged lepton diagonalization, one also expects $\theta_{13}\sim \mathcal{O}(\lambda_C)$. A value of $\theta_{13}$ near the present bound would be a strong indication in favour of this mechanism and a hint that the closeness of the measured values of the mixing angles to the TB values may be purely an accident. In addition, a very important feature of our model is that the shift of $\sin^2{\theta_{23}}$ from the maximal mixing value of 1/2 vanishes at NLO and is expected to be of $\mathcal{O}(\lambda_C^2)$ at most. In our $S_4$ model, this property is obtained by only allowing  the breaking of $S_4$ in the neutrino sector via flavons transforming as 1 and 3 (in particular with no doublets).

In order to reproduce the experimental value of the small parameter $r=\Delta m^2_{sun}/\Delta m^2_{atm}$ we need some amount of fine tuning. For instance, the RH neutrino Majorana mass $M$ should be below the cutoff $\Lambda$ (this is reminiscent of the fact that empirically $M \sim M_{GUT}$ rather than $M \sim M_{Planck}$). The neutrino spectrum is mainly of the normal hierarchy type (or moderately degenerate), the smallest light neutrino mass and the $0\nu \beta \beta$ parameter $|m_{ee}|$ are expected to be larger than about $0.1~meV$. The model is compatible with the observed amount of the baryon asymmetry in the Universe interpreted as an effect of leptogenesis.

\section*{Acknowledgements}
L.M. would like to thank the CERN Theory Division for the kind hospitality during the Fall 2008.
We recognize that this work has been partly supported by the Italian Ministero dell'Universit\`a e della Ricerca Scientifica, under the COFIN program (PRIN 2006) and by the European Commission
under contracts MRTN-CT-2006-035505 and MRTN-CT-2004-503369.
We thank Davide Meloni for some interesting comments and discussions.

\vfill
\newpage

\section*{Appendix A: the group $S_4$}

We recall here the multiplication table for $S_4$ and we list the Clebsch-Gordan coefficients in our basis. In the following we
use $\alpha_i$ to indicate the elements of the first representation of the product and $\beta_i$ to indicate those of the second representation.\\

We start with all the multiplication rules which include the 1-dimensional representations:
\[
\begin{array}{l}
1\otimes Rep=Rep\otimes1=Rep\quad\rm{with~Rep~any
~representation}\\[-10pt]
\\[8pt]
1'\otimes1'=1\sim\alpha\beta\\[-10pt]
\\[8pt]
1'\otimes2=2\sim\left(\begin{array}{c}
                    \alpha\beta_2 \\
                    -\alpha\beta_1 \\
            \end{array}\right)\\[-10pt]
\\[8pt]
1'\otimes3=3'\sim\left(\begin{array}{c}
                    \alpha\beta_1 \\
                    \alpha\beta_2 \\
                    \alpha\beta_3\\
                    \end{array}\right)\\[-10pt]
\\[8pt]
1'\otimes3'=3\sim\left(\begin{array}{c}
                            \alpha\beta_1 \\
                            \alpha\beta_2 \\
                            \alpha\beta_3\\
                    \end{array}\right)
\end{array}
\]
The multiplication rules with the 2-dimensional
representation are the following ones:
\[
\begin{array}{ll}
2\otimes2=1\oplus1'\oplus2&\quad
\rm{with}\quad\left\{\begin{array}{l}
                    1\sim\alpha_1\beta_1+\alpha_2\beta_2\\[-10pt]
                    \\[8pt]
                    1'\sim\alpha_1\beta_2-\alpha_2\beta_1\\[-10pt]
                    \\[8pt]
                    2\sim\left(\begin{array}{c}
                        \alpha_2\beta_2-\alpha_1\beta_1 \\
                        \alpha_1\beta_2+\alpha_2\beta_1\\
                    \end{array}\right)
                    \end{array}
            \right.\\[-10pt]
\\[5pt]
2\otimes3=3\oplus3^\prime&\quad
\rm{with}\quad\left\{\begin{array}{l}
                    3\sim\left(\begin{array}{c}
                        \alpha_1\beta_1\\
                        \frac{\sqrt3}{2}\alpha_2\beta_3-\frac{1}{2}\alpha_1\beta_2 \\
                        \frac{\sqrt3}{2}\alpha_2\beta_2-\frac{1}{2}\alpha_1\beta_3 \\
                    \end{array}\right)\\[-10pt]
                    \\[8pt]
                    3'\sim\left(\begin{array}{c}
                        -\alpha_2\beta_1\\
                        \frac{\sqrt3}{2}\alpha_1\beta_3+\frac{1}{2}\alpha_2\beta_2 \\
                        \frac{\sqrt3}{2}\alpha_1\beta_2+\frac{1}{2}\alpha_2\beta_3 \\
                    \end{array}\right)\\
                    \end{array}
            \right.\\[-10pt]
\\[5pt]
2\otimes3'=3\oplus3^\prime&\quad
\rm{with}\quad\left\{\begin{array}{l}
                    3\sim\left(\begin{array}{c}
                       -\alpha_2\beta_1\\
                        \frac{\sqrt3}{2}\alpha_1\beta_3+\frac{1}{2}\alpha_2\beta_2 \\
                        \frac{\sqrt3}{2}\alpha_1\beta_2+\frac{1}{2}\alpha_2\beta_3 \\
                    \end{array}\right)\\[-10pt]
                    \\[8pt]
                    3'\sim\left(\begin{array}{c}
                        \alpha_1\beta_1\\
                        \frac{\sqrt3}{2}\alpha_2\beta_3-\frac{1}{2}\alpha_1\beta_2 \\
                        \frac{\sqrt3}{2}\alpha_2\beta_2-\frac{1}{2}\alpha_1\beta_3 \\
                    \end{array}\right)\\
                    \end{array}
            \right.\\
\end{array}
\]

The multiplication rules involving the 3-dimensional
representations are:
\[
\begin{array}{ll}
3\otimes3=3'\otimes3'=1\oplus2\oplus3\oplus3'\qquad
\rm{with}\quad\left\{
\begin{array}{l}
1\sim\alpha_1\beta_1+\alpha_2\beta_3+\alpha_3\beta_2\\[-10pt]
                    \\[8pt]
2\sim\left(
     \begin{array}{c}
       \alpha_1\beta_1-\frac{1}{2}(\alpha_2\beta_3+\alpha_3\beta_2)\\
       \frac{\sqrt3}{2}(\alpha_2\beta_2+\alpha_3\beta_3)\\
     \end{array}
   \right)\\[-10pt]
   \\[8pt]
3\sim\left(\begin{array}{c}
         \alpha_3\beta_3-\alpha_2\beta_2\\
         \alpha_1\beta_3+\alpha_3\beta_1\\
         -\alpha_1\beta_2-\alpha_2\beta_1\\
        \end{array}\right)\\[-10pt]
        \\[8pt]
3'\sim\left(\begin{array}{c}
         \alpha_3\beta_2-\alpha_2\beta_3\\
         \alpha_2\beta_1-\alpha_1\beta_2\\
         \alpha_1\beta_3-\alpha_3\beta_1\\
    \end{array}\right)
\end{array}\right.\\\\[10pt]
3\otimes3'=1'\oplus2\oplus3\oplus3'\qquad
\rm{with}\quad\left\{
\begin{array}{l}
1'\sim\alpha_1\beta_1+\alpha_2\beta_3+\alpha_3\beta_2\\[-10pt]
        \\[8pt]
2\sim\left(
     \begin{array}{c}
     \frac{\sqrt3}{2}(\alpha_2\beta_2+\alpha_3\beta_3)\\
     -\alpha_1\beta_1+\frac{1}{2}(\alpha_2\beta_3+\alpha_3\beta_2)\\
     \end{array}
   \right)\\[-10pt]
        \\[8pt]
3\sim\left(\begin{array}{c}
         \alpha_3\beta_2-\alpha_2\beta_3\\
         \alpha_2\beta_1-\alpha_1\beta_2\\
         \alpha_1\beta_3-\alpha_3\beta_1\\
    \end{array}\right)\\[-10pt]
        \\[8pt]
3'\sim\left(\begin{array}{c}
         \alpha_3\beta_3-\alpha_2\beta_2\\
         \alpha_1\beta_3+\alpha_3\beta_1\\
         -\alpha_1\beta_2-\alpha_2\beta_1\\
    \end{array}\right)\\
\end{array}\right.
\end{array}
\]

\end{document}